\documentstyle[12pt,pictex,epsfig,amssymb,amsbsy]{article}

\def\appendix#1{
  \addtocounter{section}{1}
  \setcounter{equation}{0}
  \renewcommand{\thesection}{\Alph{section}}
  \section*{Appendix \thesection\protect\indent \parbox[t]{11.715cm} {#1}}
  \addcontentsline{toc}{section}{Appendix \thesection\ \ \ #1}
  }

\newcommand{\bbox}[1]{\boldsymbol{#1}}
\newcommand {\defeq}{\stackrel{\rm def}{=}}
\newcommand{\tr}[1]{\:{\rm tr}\,#1}

\def\e{{\,\rm e}\,}

\def\d{{\rm d}}

\def\i{{\rm i}}
\def\C{{x}}

\newcommand{\rf}[1]{(\ref{#1})}
\newcommand{\eq}[1]{Eq.~(\ref{#1})}
\def\be{\begin{equation}}
\def\ee{\end{equation}}
\def\beq{\begin{equation}}
\def\eeq{\end{equation}}
\def\bea{\begin{eqnarray}}
\def\eea{\end{eqnarray}}
\def\LA{\left\langle}
\def\RA{\right\rangle}
\def\A{{\mathcal{A}}}

\newcommand{\non}{\nonumber \\*}
\newcommand{\ie}{{\it i.e.}\ }

\newcommand{\ra}{\rightarrow}
\newcommand{\fr}[2]{{\textstyle {#1 \over #2}}}

\def\ga{\mathrel{\mathpalette\fun >}}
\def\fun#1#2{\lower3.6pt\vbox{\baselineskip0pt\lineskip.9pt
\ialign{$\mathsurround=0pt#1\hfil##\hfil$\crcr#2\crcr\sim\crcr}}}
\def\hybrid{\topmargin 0pt      \oddsidemargin 0pt
        \headheight 0pt \headsep 0pt
        \textwidth 17.5cm
        \textheight 25cm
        \voffset=-0.7cm
        \hoffset=-0.4cm
       \hoffset=-1.2cm
        \marginparwidth 0.0in
        \parskip 5pt plus 1pt   \jot = 1.5ex}
\catcode`\@=11
\def\marginnote#1{}

\newcount\hour
\newcount\minute
\newtoks\amorpm
\hour=\time\divide\hour by60
\minute=\time{\multiply\hour by60 \global\advance\minute by-\hour}
\edef\standardtime{{\ifnum\hour<12 \global\amorpm={am}%
        \else\global\amorpm={pm}\advance\hour by-12 \fi
        \ifnum\hour=0 \hour=12 \fi
        \number\hour:\ifnum\minute<10 0\fi\number\minute\the\amorpm}}
\edef\militarytime{\number\hour:\ifnum\minute<10 0\fi\number\minute}

\def\draftlabel#1{{\@bsphack\if@filesw {\let\thepage\relax
   \xdef\@gtempa{\write\@auxout{\string
      \newlabel{#1}{{\@currentlabel}{\thepage}}}}}\@gtempa
   \if@nobreak \ifvmode\nobreak\fi\fi\fi\@esphack}
        \gdef\@eqnlabel{#1}}
\def\@eqnlabel{}
\def\@vacuum{}
\def\draftmarginnote#1{\marginpar{\raggedright\scriptsize\tt#1}}

\def\draft{\oddsidemargin -0.1truein
        \def\@oddfoot{\sl preliminary draft \hfil
        \rm\thepage\hfil\sl\today\quad\militarytime}
        \let\@evenfoot\@oddfoot \overfullrule 3pt
        \let\label=\draftlabel
        \let\marginnote=\draftmarginnote
   \def\@eqnnum{{\rm (\theequation)}\rlap{\kern\marginparsep\tt\@eqnlabel}%
\global\let\@eqnlabel\@vacuum}  }
\newdimen\linethick  \linethick=0.4pt
\newdimen\hboxitspace    \hboxitspace=5pt
\newdimen\vboxitspace    \vboxitspace=5pt

\def\fr#1{%
\beq\new
\vcenter{
\hrule height\linethick
           \hbox{\vrule width\linethick
                 \kern\hboxitspace
                 \vbox{\kern\vboxitspace
                       \hbox{$\begin{array}{c}\displaystyle#1
          \end{array}$}%
                       \kern\vboxitspace}%
                 \kern\hboxitspace
                 \vrule width\linethick}%
           \hrule height\linethick}%
\eeq}
\newdimen\Squaresize \Squaresize=14pt
\newdimen\Thickness \Thickness=0.5pt

\def\Square#1{\hbox{\vrule width \Thickness
   \vbox to \Squaresize{\hrule height \Thickness\vss
      \hbox to \Squaresize{\hss#1\hss}
   \vss\hrule height\Thickness}
\unskip\vrule width \Thickness}
\kern-\Thickness}

\def\Vsquare#1{\vbox{\Square{$#1$}}\kern-\Thickness}

\def\numberbysection{\@addtoreset{equation}{section}
        \def\theequation{\thesection.\arabic{equation}}}
\numberbysection

\renewcommand{\theequation}{\thesection.\arabic{equation}}
\newcommand{\l@qq}[2]{\addvspace{2em}
 \hbox to\textwidth{\hspace{1em}\bf #1 \dotfill #2}}

\newcounter{app}

\def\app{\setcounter{equation}{0}
\def\theequation{\Alph{app}.\arabic{equation}}\par
   \addvspace{4ex}
   \@afterindentfalse
  \secdef\@app\@dapp}
\newcommand\@app{\@startsection {app}{1}{0ex}%
                                   {-3.5ex \@plus -1ex \@minus -.2ex}%
                                   {2.3ex \@plus.2ex}%
                                   {\normalfont\Large\bf}}

\def\@dapp#1{%
{\parindent \z@ \raggedright  \bf #1}\par\nobreak}
\def\l@app#1#2{\ifnum \c@tocdepth >\z@
    \addpenalty\@secpenalty
    \addvspace{1.0em \@plus\p@}%
    \setlength\@tempdima{2.5em}%
    \begingroup
      \parindent \z@ \rightskip \@pnumwidth
      \parfillskip -\@pnumwidth
      \leavevmode \bfseries
      \advance\leftskip\@tempdima
      \hskip -\leftskip
      #1\nobreak\hfil \nobreak\hb@xt@\@pnumwidth{\hss #2}\par
    \endgroup\fi}
\newcounter{sapp}[app]

\def\sapp{\def\theequation{\Alph{app}.\arabic{equation}}\par
   \@afterindentfalse
  \secdef\@sapp\@dsapp}
\newcommand\@sapp{\@startsection{sapp}{2}{\z@}%
                                     {-3.25ex\@plus -1ex \@minus -.2ex}%
                                     {1.5ex \@plus .2ex}%
                                     {\normalfont\large\bfseries}}

\def\@dsapp#1{%
{\parindent \z@ \raggedright  \bf #1}\par\nobreak}
\newcommand{\l@sapp}{\@dottedtocline{2}{1.5em}{3em}}

\newcounter{ssapp}[sapp]

\def\ssapp{\def\theequation{\Alph{app}.\arabic{equation}}\par
   \@afterindentfalse
  \secdef\@ssapp\@dssapp}
\newcommand\@ssapp{\@startsection{ssapp}{2}{\z@}%
                                     {-3.25ex\@plus -1ex \@minus -.2ex}%
                                     {1.5ex \@plus .2ex}%
                                     {\normalfont\normalsize\bfseries}}

\def\@dssapp#1{%
{\parindent \z@ \raggedright  \bf #1}\par\nobreak}
\newcommand{\l@ssapp}{\@dottedtocline{2}{1.5em}{3em}}

\def\titlepage{\@restonecolfalse\if@twocolumn\@restonecoltrue\onecolumn
     \else \newpage \fi \thispagestyle{empty}\c@page\z@
        \def\thefootnote{\fnsymbol{footnote}} }

\def\endtitlepage{\if@restonecol\twocolumn \else  \fi
        \def\thefootnote{\arabic{footnote}}
        \setcounter{footnote}{0}}  %\c@footnote\z@ }
\relax

\hybrid

\newtoks\@stequation

\def\subequations{\refstepcounter{equation}%
  \edef\@savedequation{\the\c@equation}%
  \@stequation=\expandafter{\theequation}%   %only want \theequation
  \edef\@savedtheequation{\the\@stequation}% %expanded once
  \edef\oldtheequation{\theequation}%
  \setcounter{equation}{0}%
  \def\theequation{\oldtheequation\alph{equation}}}

\def\endsubequations{%
  \setcounter{equation}{\@savedequation}%
  \@stequation=\expandafter{\@savedtheequation}%
  \edef\theequation{\the\@stequation}%
  \global\@ignoretrue}

\makeatletter
\newdimen\normalarrayskip              % skip between lines
\newdimen\minarrayskip                 % minimal skip between lines
\normalarrayskip\baselineskip
\minarrayskip\jot
\newif\ifold             \oldtrue            \def\new{\oldfalse}
\def\arraymode{\ifold\relax\else\displaystyle\fi} % mode of array enrties
\def\@arrayskip{\ifold\baselineskip\z@\lineskip\z@
     \else
     \baselineskip\minarrayskip\lineskip1\baselineskip\fi}

\def\@arrayclassz{\ifcase \@lastchclass \@acolampacol \or
\@ampacol \or \or \or \@addamp \or
   \@acolampacol \or \@firstampfalse \@acol \fi
\edef\@preamble{\@preamble
  \ifcase \@chnum
     \hfil$\relax\arraymode\@sharp$\hfil
     \or $\relax\arraymode\@sharp$\hfil
     \or \hfil$\relax\arraymode\@sharp$\fi}}

\makeatother
\def\bse{\begin{subequations}}                %%%SUBEQUATIONS
\def\ese{\end{subequations}}                 %
\begin{document}
\begin{titlepage}
\begin{flushright}
%NBI--HE--04--??\\
ITEP--TH--37/03\\
hep-th/0406187\\
June, 2004
\end{flushright}
\vspace{1.3cm}

\begin{center}
%{\LARGE Anomalous Behavior of Gauge Theory \\%[.4cm] 
%on a Noncommutative Plane: \\[.4cm] 
{\LARGE Wilson Loops in 2D Noncommutative  \\%[.4cm]
Euclidean Gauge Theory: \\[.4cm] 
1. Perturbative Expansion}\\
\vspace{1.4cm}
{\large Jan Ambj{\o}rn$^{a),\;c)}$, Andrey Dubin$^{b)}$ 
and Yuri Makeenko$^{a),\; b)}$}
%\footnote{E--mail: makeenko@itep.ru \ \ \ \ makeenko@nbi.dk \ } 
\\[.8cm]
%\vskip 0.2 cm
{$^{a)}$\it The Niels Bohr Institute,} \\
{\it Blegdamsvej 17, 2100 Copenhagen {\O}, Denmark}\\[.4cm]
{$^{b)}$\it Institute of Theoretical and Experimental Physics,}
\\ {\it B. Cheremushkinskaya 25, 117259 Moscow, Russia}\\[.4cm]
{$^{c)}$\it Institute for Theoretical Physics, 
Utrecht University,} \\
{\it Leuvenlaan 4, NL-3584 CE Utrecht, The Netherlands. 
}
\end{center}
\vskip 1.5 cm

\begin{abstract}

We calculate 
quantum averages of  Wilson loops (holonomies)
in gauge theories on the Euclidean noncommutative plane, using a path-integral
representation of the star-product.
We show how the perturbative expansion emerges from
a concise general formula and demonstrate its anomalous behavior
at large parameter of noncommutativity
for the simplest nonplanar diagram of genus 1.
We discuss various UV/IR regularizations of the two-dimensional
noncommutative gauge theory in the axial gauge and,
using the noncommutative loop equation,  
construct a consistent regularization.

\end{abstract}
\end{titlepage}

\newpage

\section{Introduction}

Noncommutative field theories 
attracted a lot of attention when they appeared in 
the context of M(atrix) Theory~\cite{CDS98} and
certain string models~\cite{SW99} (see \cite{Dougl&Nek} for a review and
references therein). However, the definition and study of 
these theories predates \cite{CDS98,SW99} 
and is an  interesting
subject in its own right (for a recent review see \cite{Szabo}). 
In certain sense these theories provide
us with a minimal quasi-local extension of  ordinary local field 
theories, which remains  tractable in a number
of ways. 

In short, given a commutative field theory defined in Euclidean
space ${\Bbb R}^{D}$ by the action
\be 
S=\int \d^{D}x~{\cal L}(\phi(x))\,, 
\ee
the
corresponding noncommutative theory is implemented by modifying products of
the fields $\phi({\bf x})$ to so-called star-products,
introduced according to the rule                  
\be
\left( f_{1}\ast{f_{2}}\right)({\bbox x})\equiv
\exp\left(-\frac{\i}{2}\theta_{\mu\nu}\partial^{y}_{\mu}
\partial^{z}_{\nu}\right)~
f_{1}({\bbox y})~f_{2}({\bbox z})\Bigg|_{y=z=x}.
\label{1.8}
\ee
In (\ref{1.8}) $\theta_{\mu\nu}$, the so-called parameter of noncommutativity, 
which enters the commutation relation 
\be
\left[\hat{x}_{\mu},\hat{x}_{\nu}\right]=-\i\,\theta_{\mu\nu} \hat{1} 
\label{commutator}
\ee 
is real and
antisymmetric. 

In particular, the action of the standard Yang-Mills theory is
changed to  
\be
S=\frac{1}{4g^{2}}\int \d^{D}x~
\tr\left({\mathcal{F}}^{2}_{\mu\nu}({\bbox x})\right),
\label{1.7}
\ee
where
\be
{\mathcal{F}}_{\mu\nu}=\partial_{\mu}{\mathcal{A}}_{\nu}+
\partial_{\nu}{\mathcal{A}}_{\mu}
-\i\left({\mathcal{A}}_{\mu}\ast{\mathcal{A}}_{\nu}-{\mathcal{A}}_{\nu}\ast
{\mathcal{A}}_{\mu}\right),
\label{defF}
\ee
and ${\mathcal{A}}_{\mu}\equiv{{\mathcal{A}}^{a}_{\mu}t^{a}}$ with
$\tr(t^{a}t^{b})=\delta^{ab}$. In this article we will 
only deal with the two-dimensional gauge theories and we 
have $\theta_{21}=-\theta_{12}=\theta$.

Noncommutative quantum field theories are closely related to
the twisted Eguchi--Kawai models (TEK) 
which have been known~\cite{EN83,GAK83} 
since the early 1980's. 
These models are constructed in such a way~\cite{EN83} that they in the 
large-$N$  limit reproduce the planar diagrams of 
corresponding ordinary  quantum field theories.
This relation was further pursued in Refs.~\cite{AIIKKT,ANMS99,ANMS00b}
where it was shown how noncommutative quantum field theories can be
obtained from the twisted Eguchi--Kawai models in a certain double-scaling
limit (see \cite{Mak02} for a review).

In analogy with the twisted reduced models the parameters
$\theta_{\mu\nu}$ disappear in planar diagrams of
the noncommutative quantum field theories, while  for nonplanar diagrams
it resides~\cite{Filk} in an
additional phase factor of integrands, which is determined by 
the intersection matrix.
As was argued in Ref.~\cite{Minw}, a nonplanar diagram of genus $G$
is suppressed in  $U_{\theta}(N)$ noncommutative theories at large
$\theta$ as 
\be
\frac{1}{N^{2G}\big(p^{2D} |
\det\limits_{\mu\nu} (\theta_{\mu\nu})|\big)^G} \,,
\label{suppression}
\ee
where $p$ is a typical value of external momenta.
Note that the leading orders both in $1/\theta$ and $1/N$ are
governed by the genus of the diagram.

So far, the
analysis of this and related noncommutative theories 
(with matter fields included) have been  restricted to 
a few leading orders of the perturbative expansion in $g^2$.  
Not much is known about their nonperturbative quantum dynamics,
except for the existence of certain 
classical solutions~\cite{NS98,GMS00,GN00}. 
%Secondly, the extension of program of the commutative perturbative
%renormalization has gained certain degree of success.

Before addressing the  more complicated problems of 
nonperturbative quantum dynamics,
it is reasonable to begin with a careful examination of the simplest
theory -- noncommutative (Euclidean) gauge theory in two dimensions.
The same strategy was followed in the case of ordinary gauge theories
and,  similarly to the $U(N)$ two-dimensional non-Abelian Yang--Mills 
theory~\cite{Hoo74b},
the analysis is greatly simplified by the use
of the axial gauge, where the self-interaction of the gauge field disappears.
The effects of noncommutativity are still present in the definition
of observables, for instance in 
the averages of noncommutative Wilson loops, $W(C)$,
which were introduced in Ref.~\cite{Ish} and further examined in 
Refs.~\cite{Oku99,ANMS99,ANMS00a,RU00,DR00,GHI00,LE/NCYM,RR01,DK01}.
In the  $N=\infty$ limit of ordinary  Yang--Mills theories the 
Wilson loop averages for contours without self-intersection
are given by~\cite{Kaz&Kost}
\be
\LA W(C) \RA^{(0)}_{U(N)}=\e^{-\sigma A(C)}\,, \qquad \sigma=\frac{g^2 N}2\, ,
\label{1.41b}
\ee
where $A(C)$ stands for the area of the surface enclosed
by the loop $C$. Formula (\ref{1.41b}) 
coincides with the formula obtained for an Abelian gauge theory.
In contrast to this 
the Wilson loop averages in the 2D noncommutative gauge 
theory~(\ref{1.7}) exhibit a nontrivial dependence on $\theta$. 
References 
\cite{BNT01,GSV01,BHN02,PZ02,BNT02,open1,PZ03,open2,LSZ03,DT03,BHN04}
are devoted to the analysis of 2D noncommutative gauge theory.

In the present paper we analyze 2D Euclidean noncommutative gauge 
theory perturbatively in $g^2$. We explicitly calculate 
the contribution of the nonplanar diagram of the order $g^4$ 
(having genus 1) to
the average of the noncommutative Wilson loop in ${\Bbb R}^2$:
\be
\LA W(C) \RA_{U_{\theta}(N)}
=\sum_{G=0}^{\infty}~N^{-2G}\LA W(C) \RA^{(G)}_{U_{\theta}(1)}~,
\label{CO.01}
\ee 
and find that its expansion in $1/\theta$ begins with the term 
\be
\LA W(C) \RA ^{(1)}_{U_{\theta}(1)} =
- \frac{\sigma^2}{2\pi^2}\left(1+\frac{\pi^2}3 \right) A^2(C) + 
{\cal O}(\theta^{-1}) \,.
\label{anomterm}
\ee
This anomalous term disagrees with the 
formula~\rf{suppression} and is due to the singular IR behavior
of the gauge propagator in two dimensions. 
As a consequence not only
planar diagrams survive as $\theta\ra\infty$
in the framework of the perturbative expansion of 2D noncommutative 
gauge theory.

In the companion paper~\cite{ADM04} we evaluate the contribution of
{\em all}\/ diagrams  of genus 1 to 
the noncommutative Wilson loop average~\rf{CO.01}
for a rectangular contour 
$C=\Box$ and show that at asymptotically large $\theta$ it
behaves as
\be
\LA W(\Box) \RA ^{(1)}_{U_{\theta}(1)}~\longrightarrow~
\frac{4}{\pi^2\left(\sigma\theta \right)^{2}}
\frac{\ln(\sigma{A})}{\sigma{A}}
%%\,, ~\qquad\sigma\theta \gg\sigma{A}\rightarrow{\infty}~,
\label{FA.13b}
\ee
for the areas $\sigma^{-1}\ll A(C) \ll{\theta}$ 
much larger than the string tension
$\sigma$ introduced in \eq{1.41b}, but much smaller than $\theta$.%
\footnote{In the opposite limit of $\theta/A(C)\rightarrow{0}$, the 
Wilson loop averages in two-dimensional noncommutative theory (\ref{1.7})
approaches the ones for the ordinary $U(N)$ Yang--Mills theory.}
%\be
%$\theta \gg A(C)\,$. 
%\label{LI.01}
%\ee
In particular, we find  that the perturbative and $1/\theta$ expansions of 
$\LA W(C) \RA_{U_{\theta}(N)}$ are 
{\it not}\/ interchangeable and the anomalous terms do not appear within
the $1/\theta$-expansion.

This paper is organized as follows.
In Sect.~\ref{s:2}, applying the path-integral representation~\cite{Oku99}
of the star-product, 
we derive the concise formula~(\ref{1.1}) for a generic 
Wilson loop average
$\LA W(C) \RA_{U_{\theta}(1)}$ in the noncommutative 
$U_{\theta}(1)$ gauge theory~(\ref{1.7}) and discuss how to
elevate the Abelian results to the generic case of $U_{\theta}(N)$. 
In Sect.~\ref{s:3} we apply the general formula~(\ref{1.1})
in a perturbative calculation of the noncommutative 
Wilson loop average to order $g^{4}$. 
We demonstrate the appearance of the anomalous term~\rf{anomterm}
for the simplest nonplanar diagram of genus 1 
and discuss the associated phenomenon of delocalization.
In Sect.~\ref{s:l.e.} we consider
the loop equation for the noncommutative Wilson loops in 
two-dimensional $U_{\theta}(1)$ gauge theory and use 
it to investigate their shape-(in)dependence.
In particular, we show that the anomalous term~\rf{anomterm} is
annihilated by the operator on the left-hand side of the loop equation. 
In Sect.~\ref{s:a.o.} we consider another gauge-invariant observable
which is simpler than the noncommutative Wilson loop and which
exhibits the same anomalous behavior of the perturbative expansion as
$\theta\ra\infty$. 
In Sect.~\ref{appC} we construct  
consistent UV/IR regularizations of the two-dimensional 
noncommutative gauge theory in the axial gauge,
using the noncommutative loop equation, 
and show that the Gaussian regularization is compatible with 
the usual definition of the star-product. 
Appendix~\ref{appA} is devoted to 
the derivation of the path-integral representation for the
noncommutative Wilson loops. 
Appendix~\ref{appD} contains some details of computations for the 
Gaussian regularization.
In Appendix~\ref{someu} we discuss the regularization by a finite box.

\section{Generalities\label{s:2}}

Unless otherwise specified, we will concentrate on the $D=2$ 
noncommutative $U_{\theta}(1)$ gauge theory --
the $N=1$ option of the $U_{\theta}(N)$ 
noncommutative gauge theory (\ref{1.7}),
defined on the 2D plane ${\Bbb R}^{2}$. 
The dependence on $N$ can then be restored using \eq{CO.01}.

Our aim is to analyze in this theory the average of 
{\it closed}\/ Wilson loops%
\footnote{Strictly speaking, the path-ordered 
exponential~(\ref{1.3a}) is invariant
under the noncommutative gauge transformations only after the averaging over
the gauge fields.} 
\be
W(C)={{P}} \e_{*}^{
\i \oint_{C} \d x_{\mu}(\tau) {\mathcal{A}}_{\mu}({\bbox x}(\tau))}
\label{1.3a}
\ee
defined via the star-exponential
which, as is rederived in Appendix~\ref{appA}, can be written in the 
form~\cite{Oku99}
\be
W(C)=\Bigg<\exp\bigg(\i \oint\limits_{C} \d x_{\mu}(\tau)
{\mathcal{A}}_{\mu}({\bbox x}(\tau)+{\bbox \xi}(\tau))\bigg) \Bigg>
_{\xi}~,
\label{1.3}
\ee
where the averaging over the auxiliary field $\xi_{\mu}(\tau)$ 
is to be performed according to the the path-integral representation
\be
\Bigg<{\mathcal{B}}[{\bbox \xi}(\tau)]\Bigg>_{\xi}\defeq
\int {\mathcal{D}}\xi_{\mu}(\tau)\e^{\frac{\i}{2}(\theta^{-1})_{\mu\nu}
\int \d\tau \d\tau' \xi^{\mu}(\tau)G^{-1}(\tau,\tau')
\xi^{\nu}(\tau')}~{\mathcal{B}}[{\bbox \xi}(\tau)]
\label{1.4}
\ee
with the standard flat measure. 

The smearing function $G^{-1}$ in the exponent in \eq{1.4} has 
in general
support on an interval $\varepsilon$ which plays the role of 
a regularization, and differs, as is discussed
in Appendix~\ref{appA}, from the naive one which is approached as 
$\varepsilon\ra0$: 
\beq
G^{-1}(\tau,\tau')~\stackrel{\varepsilon\ra0}{\longrightarrow}~
G^{-1}_0(\tau-\tau')=\dot \delta(\tau-\tau')~,~~~~~~
G_0(\tau-\tau')=\frac 12 {\rm sign}\,(\tau-\tau')~.
\label{Gnaive}
\eeq
But for the purposes of the present paper, it will be enough to restrict
ourselves with the case of $\varepsilon=0$ which results in the
naive form $G_0$ displayed in \eq{Gnaive}. Then we have  
\be
\Big\langle  \xi^\mu (\tau)\, \xi^\nu (\tau')
\Big\rangle = \frac{\i}{2} \theta^{\mu\nu} {\rm sign}\,(\tau-\tau')~.
\label{1.4a}
\ee

Next, in order to simplify the calculation, 
let us choose on ${\Bbb R}^{2}$ the axial
gauge
\be
{\mathcal{A}}_{1}({\bbox x})=0~\Longrightarrow~
{\mathcal{F}}_{\mu\nu}=F_{\mu\nu}\equiv
\partial_{\mu}{\mathcal{A}}_{\nu}-
\partial_{\nu}{\mathcal{A}}_{\mu}\, .
\label{1.9}
\ee
This gauge choice  reduces 
the $U_{\theta}(1)$ action (\ref{1.7}) (but {\it not}\/
the average (\ref{1.3})) to the one in the ordinary commutative $U(1)$
gauge theory. 

We can now interchange the order of averaging over $\xi$ and ${\cal A}_2$ --
\ie that over quantum fluctuations of the gauge field -- and first 
calculate the Gaussian average over ${\cal A}_2$.
As a consequence, the $\xi$-representation (\ref{1.3}) results
in the formula 
\be
\LA W(C)\RA_{U_{\theta}(1)}=\Bigg<\exp\bigg(-\frac{1}{2}
\oint\limits_{C} \d x_{\mu}(\tau)\oint\limits_{C} \d x_{\nu}(\tau')\,
D_{\mu\nu}\big({\bbox x}(\tau)-{\bbox x}(\tau')+{\bbox \xi}(\tau)-
{\bbox \xi}(\tau')\big)\bigg) \Bigg>_{\xi}~,
\label{1.1}
\ee
where $D_{\mu\nu}({\bbox z})$ is the standard propagator of the gauge field
in $D=2$, which reads in the axial gauge (\ref{1.9}) as%
\footnote{Nothing is expected to depend on the constant $B$ owing to remaining
gauge invariance.}
\be
D_{\mu\nu}({\bbox z})=\LA{\mathcal{A}}_{\mu}({\bbox z})
{\mathcal{A}}_{\nu}({\bbox 0})\RA_{U(1)}= 
{g^2}\,\delta_{\mu 2}\delta_{\nu 2}\left( B- \frac 12|z_{1}|
\right)\delta^{(1)}(z_{2})~.
\label{1.2}
\ee

In what follows we shall also need the propagators
\be
\LA F_{\mu\nu}({\bbox z})
{\mathcal{A}}_{\lambda}({\bbox 0})\RA_{U(1)}= -
\frac{g^2}{2}~ \epsilon_{\mu\nu}
\delta_{\lambda 2}\;
{\rm sign}\left(z_{1}\right)\delta^{(1)}(z_{2})
\label{1.2p}
\ee
and
\be
\LA F_{\mu\nu}({\bbox z}) F_{\rho\lambda}({\bbox 0}) \RA_{U(1)}={g^{2}}~
(\delta_{\mu \rho}\delta_{\nu \lambda}-\delta_{\mu \lambda}\delta_{\nu \rho}) 
\,\delta^{(2)}({\bbox z})~,
\label{1.11}
\ee
where $\delta^{(2)}({\bbox z})$ is the standard delta-function in $D=2$, and 
we used the fact that the propagator in the commutative $U(1)$ theory 
is given by \eq{1.2}.

It is convenient to view the variable $\tau$ in \eq{1.1} 
as an angular variable parameterizing the contour ($\tau\in{[0,2\pi]}$). 
The variable $\xi(\tau)$, over which the path integration 
is to be performed on the right-hand side of \eq{1.1}, 
depends only on the this angular variable $\tau$.
Equation \rf{1.1} contains all  information about the Wilson loops
on the noncommutative plane.
We show below how explicit formulas can be obtained starting from the
representation~\rf{1.1}. 

Note  that the average (\ref{1.1}) clearly shows the effects
of nonlocality despite the fact that 
the two-dimensional gauge theory (\ref{1.7}),
irrespectively of the value of $\theta$, has no propagating
degrees of freedom (as is manifest in axial gauge).

We also note for future reference that 
for the general $U_{\theta}(N)$ gauge theory
the string tension $\sigma$ entering Eq.~(\ref{1.41b}), resulting 
from the contribution of the planar diagrams, is
related with $g^{2}$ by the formula
\be
\sigma=g_{U_{\theta}(N)}^{2}N/2~.
\label{1.41d}
\ee

\section{Nonplanar diagrams of order $g^{4}$\label{s:3}}

In the present section we compute the first nontrivial
$g^{4}$-order of the perturbative expansion of the
average (\ref{1.1}) in $g^{2}$. For simplicity 
the contour is assumed to be non-selfintersecting. 

\subsection{Order $g^2$\label{ss:g2}}

It is instructive first to consider
the leading $g^{2}$-term.
This term is $\theta$-independent and  equal
to the corresponding contribution in the ordinary commutative $\theta=0$
Abelian $U(1)$ gauge theory. The computation is simple when 
using the 
representation (\ref{1.1}). When expanding the exponential one has
to evaluate the appropriate integral of the $\xi$-average of the
propagator
\be
\Bigg<D_{\mu\nu}({\bbox x}(\tau)-{\bbox x}(\tau')+{\bbox \xi}(\tau)-
{\bbox \xi}(\tau')) \Bigg>_{\xi}=
D_{\mu\nu}({\bbox x}(\tau)-{\bbox x}(\tau'))
\label{1.12a}
\ee
which, in fact, reduces to the propagator (\ref{1.2}) itself. 
To see this introduce the ordinary Fourier
representation of the propagator, so that the relevant $\xi$-average reads
\be
\Bigg<\e^{\i{\bbox p}\cdot({\bbox x}(\tau)-{\bbox x}(\tau')+{\bbox \xi}(\tau)-
{\bbox \xi}(\tau'))} \Bigg>_{\xi}=
\e^{\i{\bbox p}\cdot({\bbox x}(\tau)-{\bbox x}(\tau'))}\cdot
\e^{-\frac{\i}{2}\theta_{\mu\nu} p^{\mu}p^{\nu}G(\tau,\tau')}=
\e^{\i{\bbox p}\cdot({\bbox x}(\tau)-{\bbox x}(\tau'))}~,
\label{1.13}
\ee
where the last equality follows from the antisymmetry of $\theta_{\mu\nu}$.
Consequently, the $g^{2}$-order reproduces
the standard Abelian result
\be
-\frac{g^{2}}{2}\cdot A(C)~.
\label{1.13a}
\ee

It is also instructive to use the general formula
\be
\Bigg< f({\bbox \xi}(\tau_1),
{\bbox \xi}(\tau_2)) \Bigg>_{\xi}= \int
\frac{\d^{D}\xi^{\mu}_{1}\d^{D}\xi^{\mu}_{2}}{{\pi^D |\det \theta |}}
\e^{2\i \xi_1^\mu (\theta^{-1})_{\mu\nu} \xi_2^\nu}
f ({\bbox \xi}_1,{\bbox \xi}_2)
\label{product2}
\ee
for the $\xi$-average of a function that depends only on two variables
$\xi(\tau_1)$ and $\xi(\tau_2)$ (see Appendix~\ref{appA}). 
In our case $f$ depends only on the difference $\bbox \xi_1-\bbox \xi_2$,
so that, introducing the variable 
$\bbox \eta =\bbox \xi_1-\bbox \xi_2$,
we find
\bea
\int
\frac{\d^{D}\xi^{\mu}_{1}\d^{D}\xi^{\mu}_{2}}{{\pi^D |\det \theta |}}
\e^{2\i \xi_1^\mu (\theta^{-1})_{\mu\nu} \xi_2^\nu}
f ({\bbox \xi}_1 -{\bbox \xi}_2) 
&=&\int
\frac{\d^{D}\xi^{\mu}_{1}\d^{D}\eta^{\mu}}{{\pi^D |\det \theta |}}
\e^{-2\i \xi_1^\mu (\theta^{-1})_{\mu\nu} \eta^\nu}
f ({\bbox \eta}) \non
&=& \int
\d^{D}\eta^{\mu} \delta^{(D)}(\bbox \eta)
f ({\bbox \eta}) = f (\bbox 0)~,
\eea
which reproduces \eq{1.12a}.

\subsection{The order $g^4$\label{ss:g4}: planar diagram}

Turning to the next-to-leading order $g^{4}$, the 
representation~\rf{1.1} leads  to the following $\xi$-average,
\be
\Bigg<D_{22}({\bbox x}(\tau_{1})-
{\bbox x}(\tau_{3})+{\bbox \xi}(\tau_{1})-
{\bbox \xi}(\tau_{3}))~D_{22}({\bbox x}(\tau_{2})-
{\bbox x}(\tau_{4})+
{\bbox \xi}(\tau_{2})-{\bbox \xi}(\tau_{4})) \Bigg>_{\xi}.
\label{1.16}
\ee
or, after the Fourier transformation, to
\be
\Bigg<\e^{\i{\bbox p}_{1}\cdot({\bbox \xi}(\tau_{1})-{\bbox \xi}(\tau_{3}))}
\e^{\i{\bbox p}_{2}\cdot({\bbox \xi}(\tau_{2})-{\bbox \xi}(\tau_{4}))}
\Bigg>_{\xi}=
\e^{-\frac{\i}{2}\theta_{\mu\nu} q_{k}^{\mu}G_{kj}q_{j}^{\nu}}~,
\label{1.17}
\ee
where, in compliance with Eq.~(\ref{1.4a}), we have introduced
\be
G_{kj}=\frac{1}{2}~{\rm sign}\,(\tau_{k}-\tau_{j})~,
\label{1.4b}
\ee
while
\be
{\bbox q}_{1}={\bbox p}_{1}~,~~{\bbox q}_{3}=-{\bbox p}_{1}~,~~
{\bbox q}_{2}={\bbox p}_{2}~,~~{\bbox q}_{4}=-{\bbox p}_{2}~.
\label{1.18}
\ee
The ordering of ${\bbox q}_{k}$ follows the $\tau_{k}$-ordering of
${\bbox x}(\tau_{k})$, \ie the relative order of the $k$-labels
is the same as the one of the labels $\tau_{k}$ parameterizing 
%%the entire tentacle (or, equivalently, 
the position of the corresponding point
${\bbox x}(\tau_{k})$ on the loop $C$. 

\begin{figure}
\vspace*{3mm}
\centering{
\epsfig{file=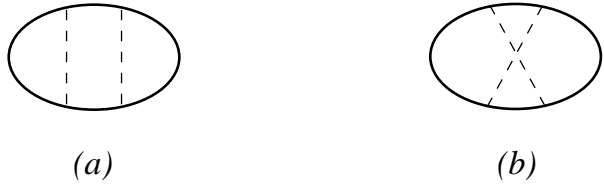}
}
\caption[]   
{Diagrams of the order $g^4$: (a) planar and (b) nonplanar. }
   \label{fi:fig.2}
\end{figure}

The formula (\ref{1.17}) actually distinguishes the
planar diagrams from the remaining nonplanar ones. 
To see this, observe
first that the topology of a particular diagram
of the perturbation-theory expansion of Eq.~(\ref{1.1}) 
is the same as the one of the
corresponding diagram in Fig.~\ref{fi:fig.2}. 
Thus, the planar configurations are
selected by the conditions
\be
\{~\tau_{1}<\tau_{2},\tau_{4}~,~~\tau_{3}>\tau_{2},\tau_{4}~\}~,~~
\{~1\leftrightarrow 2~,~~3\leftrightarrow 4~\}.
\label{1.19}
\ee
and it follows that the exponent on the
right-hand side of Eq.~(\ref{1.17}) {vanishes}
\be
\theta_{\mu\nu}~q_{k}^{\mu}~G_{kj}~q_{j}^{\nu}=0~.
\label{1.20}
\ee
The simplest way to show this is to note that the ``clusters''
$\{{\bbox q}_{1},~{\bbox  q}_{3}=-{\bbox q}_{1}\}$ and
$\{{\bbox q}_{2},~{\bbox  q}_{4}=-{\bbox  q}_{2}\}$ 
do {\it not}\/  correlate with each
other within the combination on the left-hand 
side of Eq.~(\ref{1.20}) as follows from 
\be
\sum_{j=2,4}G_{1j}~q_{j}^{\nu}=0~,~~~~~~~
\sum_{j=2,4}G_{3j}~q_{j}^{\nu}=0~,
\label{1.21}
\ee
\ie for the planar diagrams the exponent on the
right-hand side of Eq.~(\ref{1.17}) reduces to the sum of the ones
corresponding to single exchanges (\ref{1.13}) and vanishes as
well. Thus the planar part of the $g^{4}$-contribution is 
$\theta$-independent and has to be equal 
to the standard $\theta=0$ result
\be
\hbox{Fig.~\ref{fi:fig.2}a}=\frac{1}{2}\left(\frac{g^{2}}{2} A(C)\right)^{2}~.
\label{1.13b}
\ee
Together with the $g^{2}$-term (\ref{1.13a}) it provides 
the first two terms of the expansion of the well-known result
$\e^{-g^{2}A/2}$ for the loop-average associated to a non-selfintersecting
contour in the ordinary commutative $U(1)$ gauge theory.

\subsection{Order $g^4$\label{ss:g4np}: nonplanar diagram}

The situation is different for the nonplanar diagram
of  order $g^{4}$, depicted in Fig.~\ref{fi:fig.2}b. 
To be more specific consider the particular nonplanar assignment
\be
\tau_{1}<\tau_{2}<\tau_{3}<\tau_{4}~.
\label{1.22}
\ee
A straightforward computation yields
\be
\exp\left(-\frac{\i}{2}\theta_{\mu\nu}\,q_{k}^{\mu}\,G_{kj}\,q_{j}^{\nu}\right)
=
\exp\left(\i\,\theta_{\mu\nu}\,p_{1}^{\mu}\,p_{2}^{\nu}\right).
\label{1.23}
\ee
Going back to \eq{1.16}, one therefore
concludes that the average (\ref{1.16})
reproduces for the nonplanar diagram the {\it star-product}
\be
D_{22}({\bbox X})
\stackrel{\bar{\theta}}*
D_{22}({\bbox Y})=
\frac{g^4 \theta^2}{4\pi^{2}}
\frac{\exp\left(\i ({\theta}^{-1})_{\mu\nu} X_{\mu}Y_{\nu}\right)}
{X_2^2 Y_2^2}
\label{1.24}
\ee
of the two ``propagators'' (\ref{1.2}), where ${\bbox X}=
{\bbox x}(\tau_{1})-{\bbox x}(\tau_{3}),~{\bbox Y}=
{\bbox x}(\tau_{2})-{\bbox x}(\tau_{4})$ and, 
generalizing \eq{1.8}, we have obtained
\bea
f_{1}({\bbox x})\ast f_{2}({\bbox y})&=&{f_{1}({\bbox x})~
\exp\left(-\frac{\i}{2}\bar{\theta}_{\mu\nu}~
\overleftarrow{\partial^{{x}}_{\mu}}~
\overrightarrow{\partial^{{y}}_{\nu}}\right)~f_{2}({\bbox y})} \non 
& &
=\int \prod_{j=1}^{2}\frac{\d^{2}\xi^{\mu}_{j}}{({\pi^2
|\det \bar{\theta} |})^{1/2}}
\e^{2\i (\bar{\theta}^{-1})_{\mu\nu} \xi_1^\mu\xi_2 ^\nu }
f _1 ({\bbox x}+{\bbox \xi}_1)  f _2 ({\bbox y}+{\bbox \xi}_2)~.
\label{1.25}
\eea
The parameter of noncommutativity in \eq{1.24} is {\it twice}\/ larger
\be
\bar{\theta}_{\mu\nu}=2{\theta}_{\mu\nu}~,
\label{1.26}
\ee
compared to the original one ${\theta}_{\mu\nu}$ in \eq{1.4}.

Some comments concerning \eq{1.24} are in order.
If one applies $-\partial ^2 /\partial X_1^2$ to \eq{1.24}, one obtains 
\be
\delta^{(2)}({\bbox X}) 
\stackrel{\bar{\theta}}*
D_{22}({\bbox Y})=
\frac{g^2}{4\pi^{2}}
\frac{\exp\left(\i ({\theta}^{-1})_{\mu\nu} X_{\mu}Y_{\nu}\right)}
{X_2^2}.
\label{1.24fa}
\ee
Acting further by 
$-\partial ^2 /\partial Y_1^2$, one reproduces the known formula~\cite{Minw}
\be
\delta^{(2)}({\bbox X}) 
\stackrel{\bar{\theta}}*
\delta^{(2)}({\bbox Y})=
\frac{1}{4\pi^{2}\theta^2}
{\exp\left(\i ({\theta}^{-1})_{\mu\nu} X_{\mu}Y_{\nu}\right)} \,.
\label{1.24ff}
\ee

For vanishing $X_2$ or $Y_2$, the denominator on the  
right-hand side of \eq{1.24}
(or \eq{1.24fa}) is to be understood according to the prescription
\be
\frac{1}{X_2^2} \ra -\frac{\partial}{\partial X_2} 
{\cal P}\left(\frac{1}{X_2}\right),\qquad
\frac{1}{Y_2^2}\ra -\frac{\partial}{\partial Y_2} 
{\cal P}\left(\frac{1}{Y_2}\right),
\label{Pv}
\ee
where ${\cal P}$ means the principal value.%
\footnote{This prescription can be justified  by calculating
the star-product on the left-hand side of \eq{1.24} in a regularized theory
along the lines described in Sect.~\ref{appC}, 
or somewhat simpler in the present
context by simply regularizing the propagator according to 
\be
D_{22}^{\rm (R)}(X)=-\frac12 |X_1|\e^{-\mu |X_1|} 
\frac{a}{X_2^2+a^2}
~~~~{\rm or}~~~~
D_{22}^{\rm (R)}(p)= \frac{(p_1^2-\mu^2)}{(p_1^2+\mu^2)^2} \e^{-a|p_2|}.
\label{regpropp}
\ee
The IR regularization, specified by $\mu$,
is along the axis 1 and we have simultaneously introduced the UV regularization
along the axis 2 as is prescribed by the commutation relation~\rf{commutator}
which requires $a\sim\theta\mu$.
Equation~\rf{regpropp} has a nice mathematical structure 
when the regularization vanishes:
\be
D_{22}^{\rm (R)}(p)\rightarrow -
\frac{\partial}{\partial p_1}  {\cal P} \left(\frac{1}{p_1}\right) .
\label{proppc=2}
\ee  
This reproduces the prescription \rf{Pv}, when the regularization is removed.
}

Equation~\rf{1.24} exemplifies  a remarkable phenomenon of {\it long range}\/
dipole-dipole ``interactions'' 
(smeared at the scale $\sim{\sqrt{\theta}}$)
between the contour-elements 
entering into \eq{1.1}. The interactions can be 
traced back to the nonlocality of the star-product 
(\ref{1.25}), and is thus 
built into the noncommutative Wilson loop (\ref{1.3}).\footnote{More 
complicated nonplanar diagrams of perturbation theory
introduce more general multipole interactions between an arbitrary
number of ``dipoles'' made of pairs of the contour elements.} 
Conceptually this phenomenon of ``delocalization'' which is enforced
by the requirement of the noncommutative gauge invariance for the
loop-observables is noteworthy since the 2D noncommutative gauge
theory, like ordinary 2D gauge theories, lacks 
any propagating degrees of freedom. 
The quasi-locality of the interactions
is thus in sharp contrast with the short-range {\it contact}\/
interactions between the contour elements in the commutative 2D gauge
theory with a  propagator of the form \eq{1.2}. 

This phenomenon can be viewed as the generalization
of the delocalization emphasized in \cite{Minw}
for the star-product of two $\delta$-functions. 
In the context of \eq{1.24} the arguments of \cite{Minw} can be
 applied for ${\bbox X}=
{\bbox Y}$ when the right-hand side  of \eq{1.24ff} becomes
constant that, in turn, refers to the infinite range of the ``quasi-locality''.
The latter infinity precisely matches the $\Delta\rightarrow{0}$ option
of the estimate
\be
\delta\sim{{\rm max}\,\left(\Delta~,~\frac{\theta}{\Delta}\right)}
\label{1.8a}
\ee
of the characteristic ``width'' $\delta$ of the star-product (\ref{1.8}) when
the function $f_{1}(\bbox{x})=f_{2}(\bbox{x})=f(\bbox{x})$  itself has 
a ``width'' of order  $\Delta$.

\subsection{Nonplanar order $g^4$\label{ss:g4npc}(continued): 
anomalous behavior}

Next, according to the formula~\rf{1.1},  the right-hand
side of \eq{1.24} is to be integrated
over those positions of ${\bbox x}(\tau_{k}),~k=1,...,4$ 
on the contour $C$, which are consistent with the
nonplanar topology of the associated diagram. 
For the nonplanar diagram in Fig.~\ref{fi:fig.2}b, we have
\be
\hbox{Fig.~\ref{fi:fig.2}b}~=g^4 
\bbox{P}\!
\int\limits_x\!\!\!\int\limits_y\!\!\!\int\limits_z \!\!\!\int\limits_t
D_{22}(X) * D_{22}(Y)\,,
\label{fourthorder}
\ee
where explicitly
\be
\bbox{P}\!
\int\limits_x\!\!\!\int\limits_y\!\!\!\int\limits_z \!\!\!\int\limits_t
\cdots~\defeq\int_0^{2\pi} \d \tau_x
\int_{\tau_x}^{2\pi} \d \tau_y
\int_{\tau_y}^{2\pi} \d \tau_z\int_{\tau_z}^{2\pi} \d \tau_t \;
\dot \C_2({\tau_x})\dot \C_2({\tau_y})\dot \C_2({\tau_z})\dot \C_2({\tau_t})
\cdots
\label{deforedred}
\ee
with $0\leq\tau<2\pi$ parametrizing the loop $C$, given by the function
$\C_\mu(\tau)$ \mbox{($\C_\mu(0)=\C_\mu(2\pi)$),} and
we have introduced
\bea
&&x\equiv x(\tau_1)\,, \qquad y\equiv x(\tau_2)\,, \non
&&z\equiv x(\tau_3)\,, \qquad t\equiv x(\tau_4)\,, 
\eea
so that
\be
X=x-z\,,\qquad Y=y-t \,.
\ee

Generically, the
$\theta$-dependence of the resulting expression does {\it not\/} possess any
apparent topological interpretation. 
Moreover, the expansion of the right-hand side of \eq{1.24} in $1/\theta$
starts from the term $\theta^2$ rather than $1/\theta^2$ as one might have
expected. However, the 
$\theta^2$-term and the $\theta^1$-term 
of the expansion of \rf{fourthorder} in $1/\theta$
can easily be shown to vanish in accordance with \cite{DT03}.

The $\theta^0$-term of \eq{fourthorder} reads  
\be
\theta^0 \hbox{-term}~=- \frac{g^4}{8 \pi^2}\left( 
\bbox{P}\!
\int\limits_x\!\!\!\int\limits_y\!\!\!\int\limits_z \!\!\!\int\limits_t +
\bbox{P} \!
\int\limits_y\!\!\!\int\limits_x\!\!\!\int\limits_t\!\!\! \int\limits_z \right)
\left(\frac{X_1^2}{X_2^2}-\frac{X_1Y_1}{X_2 Y_2}\right) \,.
\label{theta0}
\ee 

The calculation of the $\theta^0$-term given by \rf{theta0} can be 
performed as follows.
We first integrate   ${X_1^2}/{X_2^2}$ over $y$ and $t$ to obtain
\be
%&&
\left( 
\bbox{P}\!
\int\limits_x\!\!\!\int\limits_y\!\!\!\int\limits_z \!\!\!\int\limits_t +
\bbox{P} \!
\int\limits_y\!\!\!\int\limits_x\!\!\!\int\limits_t\!\!\! \int\limits_z \right)
\frac{X_1^2}{X_2^2}
= -\bbox{P}\!
\int\limits_x\!\!\!\int\limits_z X_1^2 %\non &&
= 2 \bbox{P}\!
\int\limits_x\!\!\!\int\limits_z x_1 z_1 = \oint\limits_x x_1
\oint\limits_z z_1 = A^2
\label{theta21}
\ee
independently of the form of the contour. For the second term we use the 
Stokes theorem 
\be
\oint\limits_C \d x_2\;f(x)=\int\limits_{S(C)} \d\sigma_{12}(x) \partial_1 f(x)
\ee
for the integrals over $x$ and $y$ and obtain
\bea
-2\bbox{P}\!
 \int\limits_x\!\!\!\int\limits_y\!\!\!\int\limits_z \!\!\!\int\limits_t
\frac{X_1Y_1}{X_2 Y_2}&=& -2 
\int\limits_S \d\sigma_{12}(x)\int\limits_S\d\sigma_{12}(y) 
\int\limits_{y_2}^{x_2} \d z_2 \int\limits_{z_2}^{x_2} \d t_2 \,
\frac{1}{(x_2-z_2)(y_2-t_2)} \non &= & -2 A^2 \left(-\frac{\pi^2}6\right)
=\frac{\pi^2}3 A^2,  
\label{theta22}
\eea 
which adds with  \rf{theta21} to  
\be
\theta^0 \hbox{-term}~=- \frac{g^4}{8 \pi^2}\left(1+ \frac{\pi^2}3 
\right) A^2 \,.
\label{theta0final}
\ee

The $\theta^{-1}$-term in the expansion of \eq{1.24} is
\be
\theta^{-1} \hbox{-term}~=-\frac{\i g^4}{3!\cdot4\pi^2\theta } \left( 
\bbox{P}\!
\int\limits_x\!\!\!\int\limits_y\!\!\!\int\limits_z \!\!\!\int\limits_t -
\bbox{P} \!
\int\limits_y\!\!\!\int\limits_x\!\!\!\int\limits_t\!\!\! \int\limits_z \right)
\left(\frac{X_1^3 Y_2}{X_2^2} - 3\frac{X_1^2 Y_1}{X_2}\right)\,.
\label{theta-1}
\ee 

The explicit calculation of the right-hand side of \eq{theta-1} is as follows.
The contour integral of the first term vanishes owing to
\be
\int_{x_2}^{z_2} \d y_2 \int_{z_2}^{x_2} \d t_2 \;Y_2=0 \,.
\label{===0}
\ee
The contribution of the second term involves
\bea
\bbox{P}\!
\int\limits_x\!\!\!\int\limits_y\!\!\!\int\limits_z \!\!\!\int\limits_t
\frac{X_1^2 Y_1}{X_2} &=& \frac14 \oint\limits_C \d x_2 
\int\limits_{C_{xx}} \d z_2
\frac{X_1^2}{X_2}\int\limits_{C_{xz}} \d y_2\int\limits_{C_{zx}} \d t_2\;
(y_1-t_1) \non &=&
\frac14 \oint\limits_C \d x_2 \int\limits_{C_{xx}} \d z_2 X_1^2 \left(~
\int\limits_{C_{xz}} \d y_2 \,y_1 +\int\limits_{C_{zx}} \d t_2\,t_1\right)
\non &=& -\frac 12 \oint\limits_x x_1\oint\limits_z z_1\oint\limits_y y_1
=-\frac 12 A^3 \,.
\eea
We thus obtain
\be
\theta^{-1} \hbox{-term}~=-\frac{\i g^4 A^3}{8\pi^2\theta }
\label{pureimagine}
\ee
which is pure imaginary. The sign depends on the orientation of the contour.

The $\theta^{-2}$-term is given by the expression
\be
\theta^{-2} \hbox{-term}~=\frac{g^4}{4!\cdot 4\pi^2\theta^2 } \left( 
\bbox{P}\!
\int\limits_x\!\!\!\int\limits_y\!\!\!\int\limits_z \!\!\!\int\limits_t +
\bbox{P} \!
\int\limits_y\!\!\!\int\limits_x\!\!\!\int\limits_t\!\!\! \int\limits_z \right)
\left(\frac{X_1^4 Y_2^2}{X_2^2} - 4\frac{X_1^3 Y_1 Y_2}{X_2}
+3 X_1^2 Y_1^2\right)\,.
\label{theta-2}
\ee 

We have not attempted a contour independent 
calculation of \rf{theta-2}. However, for a circle, 
the difference of the two
contour integrals in \eq{theta-2} can be calculated.
The result is
\be
R^4\left(\pi^4 + \frac{175 \pi^2}{12} \right) = A
^4\left(1+\frac{175}{ 12\pi^2} \right),
\label{order-2circle}
\ee
where $R$ is the radius and $A$ is the area.
This result (as well as \rf{pureimagine}) agrees with
those of Bassetto et al.~\cite{BNT01,BNT02} 
for the Wu--Mandelstam--Leibbrandt propagator
in Minkowski space. 
The coefficient in the $\theta^0$-term differs. However, it agrees 
with the result obtained by the same authors using the principal 
value presciption for the propagator \cite{BNT01}.\footnote{We 
thank A. Bassetto, A. Torrielli and F. Vian for pointing this out to us.}

However, for a rectangle it is straightforward to perform the 
integrals in \rf{theta-2} and we obtain:
\be
A^4\left( \frac1{18}+ 1 + \frac 32\right)= \frac{23}9 A^4\,.
\label{order-2recta}
\ee
The conclusion is that the $\theta^{-2}$ contribution to the 
Wilson loop average is {\em not}\/ shape independent.\footnote{Again we 
would like to thank A. Bassetto, A. Torrielli and F. Vian 
for communicating to us that the result  
\rf{order-2circle} is invariant under deformations of the circle
to an ellipse.}

\section{The NC loop equation\label{s:l.e.}}

The noncommutative loop equation~\cite{LE/NCYM}  
%%(Eq.~(16.73) in my book):
\be
\epsilon^{\mu\nu}\partial_\nu^x \frac{\delta}{\delta\sigma_{12}(x)}
\LA{\cal W}_{\rm clos}(C)  \RA_{\cal A}= 
\frac{g^2 {\cal V}}{(2\pi)^2 \theta^2} 
\oint\limits_C \d z^\mu 
\LA{\cal W}_{\rm open}(C_{xz})\,{\cal W}_{\rm open}(C_{zx} ) \RA_{\cal A} 
\label{NCle}
\ee
relates the average of the closed
Wilson loop to the correlator of two open Wilson loops given in the axial
gauge by
\be
{\cal W}_{\rm open}(C_{0\eta})= \int\limits_V \d^2 u
\e_*^{\i \int _{C_{u(u+\eta)}}\d \C_2 A_2(u+\C)} \e^{\i\eta \wedge u/\theta}\,,
\label{Wopenmod}
\ee 
where $\eta \wedge u\equiv \eta_1 u_2 - \eta_2 u_1$.
Following Ref.~\cite{Mak02}, we have introduced in \eq{NCle}
a unit volume ${\cal V}$ (${\cal V}=1$ for a box).
The planar contribution comes from the factorized part of the 
correlator (the first term on the right-hand side of)
\be
\LA{\cal W}_{\rm open}(C_{xz})\,{\cal W}_{\rm open}(C_{zx} ) \RA =
\LA{\cal W}_{\rm open}(C_{xz})\RA \LA{\cal W}_{\rm open}(C_{zx} ) \RA 
+\LA{\cal W}_{\rm open}(C_{xz})\,{\cal W}_{\rm open}(C_{zx} ) \RA_{\rm conn} 
\label{correlat}
\ee
which is proportional to the (smeared) $\delta$-function
$\delta^{(2)}(x-z)$.

This is because
\be
\LA{\cal W}_{\rm open}(C_{xz}) \RA_A \propto
\int\limits_V \d^2 u \e^{\i(x-z)^\lambda \theta^{-1}_{\lambda\nu} u^\nu}
=(2 \pi)^2 \delta_{\mu}^{(2)}\left((x-z)\theta^{-1}\right).
\label{o0}
\ee
The precise form of $\delta_{\mu}^{(2)}$ depends on the
IR regularization. For a Gaussian spherically symmetric IR cutoff we have
\be
\int\limits_V \d^2 u \cdots = \int \d^2 u \e^{-\mu^2 u^2/2} \cdots \,,
\qquad V=\frac{2\pi}{\mu^2}
\ee
and 
\be
\delta_{\mu}^{(2)}(\eta\theta^{-1})=\frac{1}{2\pi  \mu^2}
\e^{-\eta^2/2\theta^2\mu^2}.
\label{deltaspere}
\ee
For a box, which possesses only cubic symmetry, we have
\be
\delta_{\mu}^{(2)}(\eta\theta^{-1})=
\prod\limits_{i=1}^2\frac{\theta}{\pi\eta_i}
\sin\frac{\eta_i}{\theta\mu}\,, \qquad V=\frac{4}{\mu^2}\,.
\label{deltabox}
\ee

The standard $\delta$-function in the factorized term 
on the right-hand side of the 
loop equation is reproduced for $\mu\ra0$ as%
\footnote{The volume element ${\cal V}$ in \eq{NCle} is equal to 1 for the cube
and to $2^{d/2}$ for the spherical regularization in $d$-dimensions.}  
\be
\frac {\cal V}{(2\pi)^2\theta^2}
\left[(2\pi)^2 \delta_{\mu}^{(2)}(\eta\theta^{-1})\right]^2 =
V \delta_{\theta\mu/\sqrt{2}}^{(2)}(\eta)
\label{smeasphere}
\ee
for the sphere or
\be
\frac1{(2\pi)^2 \theta^2} 
\left[(2\pi)^2\prod\limits_{i=1}^2 \frac{\theta}{\pi \eta_i} 
\sin \frac{\eta_i}{\theta\mu} \right]^2= V
\prod\limits_{i=1}^2 \frac{\theta\mu}{\pi \eta^2_i} 
\sin^2 \frac{\eta_i}{\theta\mu} \ra  V\delta^{(2)}(\eta)
\label{smearcube}
\ee
for the cube. Therefore, the factorized part of the correlator~\rf{correlat}
reproduces the loop equation of the $N=\infty$ Yang--Mills theory.

Alternatively, the contribution of the connected correlator in \eq{correlat}
to the right-hand side of the NC loop equation~\rf{NCle} is suppressed
at large $\theta$ as $1/\theta^2$, so that \eq{NCle} reproduces 
the loop equation of 
the $N=\infty$ Yang--Mills theory as $\theta\ra\infty$.
The expectation that only planar diagrams survive as $\theta\ra\infty$
is based, in particular, on this argument.  

\subsection{Anomalous terms as zero modes}

A question which immediately arises is why the existence of the anomalous
term~\rf{theta0final} (or \rf{pureimagine}) 
does not contradict the arguments of the previous
paragraph. We show in this subsection that the anomalous terms are
zero modes of the operator on the left-hand side of \eq{NCle}.

Let us first verify how the gauge-invariant loop equation is satisfied in
the axial gauge in 2D. To the order $g^4$ we have for the connected correlator
of the two open Wilson loops on the right-hand side:
\bea
%&&
\lefteqn{-\frac{g^4{\cal V}}{4\pi^2\theta^2} \oint\limits_C \d z_\nu
\int\limits_{C_{xz}} \d y_2 
\int\limits_{C_{zx}} \d t_2 \int\limits_V \d^2 u\int\limits_V \d^2 v
D_{22}^{\rm (R)}(u+y-v-t) \e^{\i (x-z)^\mu\theta^{-1}_{\mu\lambda}
(u-v)^\lambda}} \non
&&= -\frac{g^4}{4\pi^2}V \oint\limits_C \d z_\nu
\int\limits_{C_{xz}} \d y_2 
\int\limits_{C_{zx}} \d t_2 D_{22}^{\rm (R)}(\theta^{-1}X_2,-\theta^{-1}X_1)
\e^{\i X\wedge Y/\theta}\,,~~~~~~
\label{4RHS}
\eea
where the propagator in the first line is in coordinate space,
and that in the second line is in momentum space.

One obtains the same result by acting with the loop operator on the
nonplanar diagram in Fig.~\ref{fi:fig.2}b as can be seen 
by applying the area derivative
to the ordered exponential:
\be
\frac{\delta}{\delta\sigma_{12}(x)} \LA \e_*^{\i \int A_2} \RA_{A_2}=
\i \LA F_{12}(x) \e_*^{\i \int A_2} \RA_{A_2} \,,
\label{areaderivative}
\ee
where $F_{12}=\partial_1 A_2$ in the axial gauge.\footnote{As 
usual when dealing with the loop equation 
we assume here that the variation of the contour is
much smaller than the UV cutoff ($\sim \theta \mu$).}
Differentiating \eq{areaderivative} with respect to $x_1$, 
expanding to the order $A^3_2$ and using \eq{1.24}, 
we obtain to the order $g^4$
\be
\LA \partial_1^x F_{12}(x) \e_*^{\i \int {A_2}} \RA_{A_2} =
-\frac{g^4}{4\pi^2} \bbox{P}\!
\int\limits_y\!\!\!\int\limits_z \!\!\!\int\limits_t
\frac{\e^{\i X\wedge Y/\theta}}{X_2^2}\,,
\label{shapeindependence}
\ee
where the ordering goes from $x$ to $x$ along the contour.
This reproduces the $\nu=2$ component of the right-hand side of \eq{4RHS}.
Differentiating \eq{areaderivative} analogously with respect to $x_2$:
\be
-\partial_2^x\frac{\delta}{\delta\sigma_{12}(x)} 
\LA \e_*^{\i \int {A_2}} \RA_{A_2}=
-\i \LA \partial_2^x F_{12}(x) \e_*^{\i \int {A_2}} \RA_{A_2}
\label{shapeindependence2}
\ee
and integrating by parts 
(the contact terms are mutually canceled), we obtain 
the $\nu=1$ component of the right-hand side of
\eq{4RHS}.

In order to calculate the result of applying 
 the loop operator to the anomalous terms, 
we note the following. To 
order $\theta^0$, \eq{shapeindependence} is reduced to
\be
\bbox{P}\!
\int\limits_y\!\!\!\int\limits_z \!\!\!\int\limits_t
\frac{1}{X_2^2} = - \oint\limits_z =0
\ee
which proves that the $\theta^0$-term~\rf{theta0} is annihilated by the 
operator on the left-hand side of the loop equation.
The same is true for the $\theta^{-1}$-term~\rf{theta-1},
whose contribution to \eq{shapeindependence} is
\be
\i \frac{g^4}{4\pi^2 \theta} \bbox{P}\!
\int\limits_y\!\!\!\int\limits_z \!\!\!\int\limits_t
\left( \frac{X_1 Y_2}{X_2^2} - \frac{Y_1}{X_2}\right).
\label{shapeindependence-1}
\ee
The integral of the first term vanishes as in \eq{===0}.
Integrating the second term,
we find
\be
\int\limits_y\int\limits_z y_1 +\int\limits_z\int\limits_t t_1 
=\oint\limits_z \oint\limits_y y_1 =0 \cdot A=0.
\ee
The same statements can be made for 
the $\theta^0$ and $\theta^{-1}$ terms of 
\eq{shapeindependence2}, given by the $\nu=1$ component 
of the right-hand side of \eq{4RHS}.

We have thus shown that the anomalous $\theta^0$- and  
$\theta^{-1}$-terms, \rf{theta0} and \rf{theta-1}, are annihilated
by the loop operator and do not 
contribute to the NC loop equation~\rf{NCle}.

\subsection{Symplectic invariance and shape-independence\label{s:sympl}}

The Wilson loop averages in the ordinary 2D Yang--Mills theory depend only on
the area enclosed by the loop. It is a consequence of symplectic invariance.
If the same holds in the noncommutative case, the area derivative
\rf{areaderivative} should not depend on
where the point $x$ is chosen on the contour.
Thus symplectic invariance implies
\be
\dot x_\mu \partial_\mu^x
\frac{\delta}{\delta\sigma_{12}(x)} \LA \e_*^{\i \int {A_2}} \RA_{A_2}=
\i \LA \dot x_\mu \partial_\mu^x F_{12}(x) 
\e_*^{\i \int {A_2}} \RA_{A_2}=0\, ,
\label{areanodep}
\ee
or, using \eq{NCle},
\be
\varepsilon_{\mu\nu}\dot x_\mu \oint\limits_C \d z^\nu 
\LA{\cal W}_{\rm open}(C_{xz})\,{\cal W}_{\rm open}(C_{zx} ) \RA_{A_2} =0\,.
\label{noshape}
\ee

For non-selfintersecting contours,
the factorized part on the right-hand side 
of \eq{correlat} always obeys \eq{noshape}
just as in the commutative case.
For the contribution of the nonplanar diagram of  order $g^4$,
depicted in Fig.~\ref{fi:fig.2}b, 
we have from \eq{4RHS}
\be
\varepsilon_{\mu\nu}\dot x_\mu \oint\limits_C \d z_\nu
\int\limits_{C_{xz}} \d y_2 
\int\limits_{C_{zx}} \d t_2  
\e^{\i X\wedge Y/\theta} \frac{\partial}{\partial X_2}
{\cal P}\frac{1}{X_2}\,.
\label{sind}
\ee

It can be shown that \rf{sind} vanishes to the orders $\theta^0$ and
$\theta^{-1}$, which agrees with what is shown in 
Subsect.~\ref{ss:g4npc}. Rather surprisingly, it also vanishes to the 
order $\theta^{-2}$ for the circle. However, it does
not vanish for the rectangle as explicit calculations show, in agreement
with the conclusion reached in Subsect.~\ref{ss:g4npc}, namely that 
the $\theta^{-2}$-terms are not shape independent.

\section{Another observable\label{s:a.o.}}

The simplest observable for which the discovered anomaly on ${\Bbb R}^2$
shows up is the average of two field strengths located at the points
$x$ and $x+\eta$:
\be
G(\eta)= \frac{1}V \LA\int \d^2 x \,{\cal F}_{12}(x) * 
\e^{\i \int _x^{x+\eta} {\cal A} }_* * {\cal F}_{12}(x+\eta) *
\e^{\i \int _{x+\eta}^x {\cal A}}_*\RA_{\cal A} , 
\label{dumbbell}
\ee
where the noncommutative phase factors (along certain paths
connecting the points $x$ and $x+\eta$) are required by the star-gauge
invariance. When these paths are chosen to be straight, the associated
contour is dumbbell shaped. 

\begin{figure}
\vspace*{3mm}
\centering{
\epsfig{file=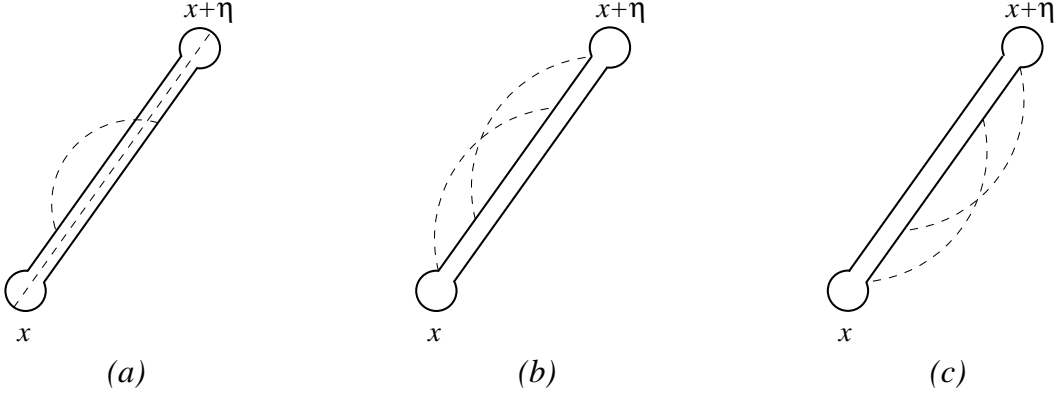}
}
\caption[]   
{Nonplanar diagrams of the order $g^4$ for $G(\eta)$ defined by \eq{dumbbella}.
The circles at the points $x$ and $x+\eta$
are associated with the field strengths $F_{12}(x)$ and $F_{12}(x+\eta)$,
respectively. The noncommutative phase factors along the straight paths
connecting them are depicted by the solid lines. 
The diagrams involve: (a) propagators $\LA F_{12}F_{12} \RA$ and
$\LA A_2 A_2 \RA$, (b) and (c) two propagators $\LA F_{12}A_{2} \RA$,
depicted by the dashed lines. }
   \label{fi:anobs}
\end{figure}

In the axial gauge, where \eq{dumbbell} takes the form
\be
G(\eta)= \frac{1}V \LA\int \d^2 x \,F_{12}(x) * 
\e^{\i \int _x^{x+\eta} {A_2} }_* * F_{12}(x+\eta) *
\e^{\i \int _{x+\eta}^x {A_2}}_*\RA_{A_2} , 
\label{dumbbella}
\ee
there are two types of nonplanar diagrams of order
$g^4$ as is depicted in Fig.~\ref{fi:anobs}. 
They involve: (a) propagators $\LA F_{12}F_{12} \RA$ and
$\LA A_2 A_2 \RA$, (b) and (c) two propagators $\LA F_{12}A_{2} \RA$, 
respectively. 

The diagram in Fig.~\ref{fi:anobs}a with $\eta=-X$ contributes 
\be
g^4\frac{\partial^2}{\partial X_1^2} \int\limits_{C_{xz}} \d y_2
\int\limits_{C_{zx}} \d t_2 \, D_{22}(X) \ast D_{22}(Y) =
- \frac{g^4}{4\pi^2} \frac1{X_2^2}\int\limits_{C_{xz}} \d y_2
\int\limits_{C_{zx}} \d t_2 \, \e^{\i X \wedge Y/\theta}.
\label{1)}
\ee
Similarly the diagrams in Fig.~\ref{fi:anobs}b with $\eta=y-x$ gives
\be
g^4\frac{\partial^2}{\partial X_1 \partial Y_1} \int\limits_{C_{yx}} \d z_2
\int\limits_{C_{zx}} \d t_2 \, D_{22}(X) * D_{22}(Y) =
\frac{g^4}{4\pi^2} \int\limits_{C_{yx}} \d z_2
\int\limits_{C_{zx}} \d t_2 \, \frac1{X_2 Y_2} 
\e^{\i X\wedge Y/\theta} 
\label{2)}
\ee
and the same contribution comes from the diagram in Fig.~\ref{fi:anobs}c
with $\eta=t-x$.

To the order $\theta^0$ we find from Eqs.~\rf{1)} and \rf{2)}
\be
G(\eta)=  \frac{g^4}{4\pi^2} \left(1+\frac{\pi^2}3\right)
\ee
%with the same coefficient as before. 
which does not depend on the form of the paths connecting 
the points $x$ and $x+\eta$.
This formula is in agreement with \eq{theta0final} and can be obtained
by acting with 
\mbox{$-\delta^2/\delta\sigma_{12}(x)\delta\sigma_{12}(x+\eta)$}
on  \eq{theta0final}, according to \eq{areaderivative}, 
which is the same as acting with $-\partial^2/\partial A^2$
since the dependence is only on the area.

For a straight path it is convenient to use  
the variables $Y_2=y_2-t_2$ and 
\be
s_2=\frac{y_2+t_2}2-x_2\,, \qquad s_2\in [\,0,X_2]
\label{newvariable}
\ee
when integrating over $y_2$ and $t_2$ in \eq{1)}
(and similarly in \eq{2)}). We then find
\be
G(\eta)=  \frac{g^4}{4\pi^2} \left(1+\frac{\pi^2}3 \right)
\label{notatall}
\ee
%which does not at all depend  on $\theta$. 
to all orders in $\theta^{-1}$.
%Some details of the derivation of \eq{notatall} for the Gaussian 
%regularization are presented in Appendix~\ref{appD}.

Equation \rf{notatall} is obtained using the ``naive'' \eq{1.24}
which is not applicable, as is already mentioned, for vanishing $X_2$ or
$Y_2$, when 
an uncertainty of the type $0\times\infty$ appears in the contour
integral. We investigate this issue in Appendix~\ref{appD}, where 
some details of the calculation are presented  
for the Gaussian regularization introduced in the next Section.
The conclusion is that \eq{notatall} does not change.

\section{Consistency of star-product with regularization\label{appC}}

As is already mentioned, the ``naive'' expression~\rf{1.24} for
the star-product of two propagators is to be regularized.
In general, the regularized expression would be
regularization-dependent.
It is slightly non-trivial to introduce a 
regularization which preserves what we understand as 
star-gauge invariance   since the star-product mixes 
the IR and UV sectors of the theory. In this Section and Appendix~\ref{appD}
we construct a possible consistent regularization, which is 
compatible with the NC loop equation and show how the ``naive'' value
given by \eq{1.24} is recovered when the cutoff is removed.

Consistent UV and IR regularizations can be constructed using the 
noncommutative loop equation~\rf{NCle}. Introducing an IR cutoff
as is described in Sect.~\ref{s:l.e.}, we get simultaneously
a UV smearing of the delta-function on the right-hand side given by 
Eqs.~\rf{smeasphere} or \rf{smearcube} for the spherically or cubic
symmetric IR cutoffs, respectively.

The strategy is thus to introduce an IR cutoff by modifying
the integral in the definition of the (open) Wilson loop~\rf{Wopenmod}
that gives \eq{o0} to the order $g^0$. As is show in Sect.~\ref{s:l.e.},
this fixes the smearing of the $\delta$-function that appears on the
right-hand side of the 
NC loop equation to the order $g^2$, which in turn fixes the
UV regularization of the propagator. This is of course a manifestation
of the usual UV/IR mixing in noncommutative theories.

The regularized propagator is then given by the convolution
\be
D^{\rm (R)}_{22}(x)= \int \d^2 z \, \delta^{(2)}_{\rm (R)}(z) D_{22}^{(0)}(x-z)
\label{DR}
\ee
which obeys
\be
-\partial_1^2 D^{\rm (R)}_{22}(x)=\delta^{(2)}_{\rm (R)}(x) \,.
\ee

Given the propagator~\rf{DR}, the right-hand side of the loop equation involves
to the order $g^4$ the correlator of two open Wilson loops
given by the first line in \eq{4RHS} which should be equal to the result of
acting by the loop operator on the nonplanar diagram of the order $g^4$
with two crossed propagator lines. 
We thus find the following formula is to be valid:
\be
V \delta^{(2)}_{\rm (R)}(X) * D^{\rm (R)}_{22}(Y)= 
\frac{{\cal V}}{4\pi^2\theta^2}
\int\limits_V \d^2 u \int\limits_V \d^2 v \, 
D^{\rm (R)}_{22}(u-v+Y) \e^{\i X\wedge (u-v) /\theta} 
\label{f11}
\ee
or, applying $-\partial^2/\partial Y_1^2$,
\be
V \delta^{(2)}_{\rm (R)}(X) * \delta^{(2)}_{\rm (R)}(Y)= 
\frac{{\cal V}}{4\pi^2\theta^2}
\int\limits_V \d^2 u\int\limits_V \d^2 v \, 
\delta^{(2)}_{\rm (R)}(u-v+Y) \e^{\i X\wedge (u-v) /\theta} 
\label{f12}
\ee
as a consequence.

Separating the volume-factor,
we thus find the following formula is to be valid for the Gaussian 
regularization:
\be
\delta^{(2)}_{\rm (R)}(X) * D^{\rm (R)}_{22}(Y)= \frac{1}{4\pi^2\theta^2}
\int\limits \d^2 u \, \e^{-\mu^2 u^2/4} 
D^{\rm (R)}_{22}(u-Y) \e^{\i X\wedge u /\theta} 
\label{f1}
\ee
or, applying $-\partial^2/\partial Y_1^2$,
\be
\delta^{(2)}_{\rm (R)}(X) * \delta^{(2)}_{\rm (R)}(Y)= \frac{1}{4\pi^2\theta^2}
\int\limits \d^2 u \, \e^{-\mu^2 u^2/4} 
\delta^{(2)}_{{\rm (R)}}(u-Y) \e^{\i X\wedge u /\theta} 
\label{f2}
\ee
as a consequence. 

The star-product on the left-hand sides of Eqs.~\rf{f1} and \rf{f2} should be
defined in a way for these formulas to be true. This is 
a consistency of the star-product in the regularized theory
with the regularization.

Making the Fourier transformation,
it is easy to see that \eq{f2} is identically satisfied for 
the Gaussian regularization when%
\footnote{It is explicitly seen from this formula that $\theta\mu$
plays the role of the UV cutoff $a$ as is prescribed by the UV/IR mixing.}
\be
\delta^{(2)}_{{\rm (R)}}(X)=\frac{1}{\pi \theta^2 \mu^2} 
\e^{-X^2 /(\theta\mu)^2 }
\label{deltaG}
\ee
by the
usual definition of the star-product 
\be
\e^{\i p X}* \e^{\i q Y}= \e^{\i p X + \i q Y} \e^{\i p \wedge q \,\theta} \,.
\ee
Similarly, \eq{f1} is satisfied for an arbitrary function 
$D^{\rm (R)}_{22}(X)$ because it is linear.

Given this definition of the star-product, we find
for the star-product of the regularized propagators:
\bea
\lefteqn
{D_{22}^{\rm (R)}(X)* D_{22}^{\rm (R)}(Y)
= 
\frac{g^4}{4\pi^2\theta^2} }\non
&&\times\int\limits_{-\infty}^{+\infty} \d u 
\e^{-\mu^2u^2/4 + \i u Y_2/\theta} \Big( B-\frac12 |u-X_1| \Big)
\int\limits_{-\infty}^{+\infty} \d v \e^{-\mu^2v^2/4 - 
\i v  X_2/\theta} \Big( B-\frac12 |v-Y_1| \Big)
\non &&
\label{starpropagatorGa}
\eea
which regularizes the ``naive'' 
\eq{1.24} for $X_2,Y_2\lesssim \mu\theta$, reproducing the usual
product of the two propagators~\rf{1.2} as $\theta\ra0$.

%\section{Conclusions\label{s:8}}

\subsection*{Acknowledgments}
Warm thanks to Antonio Bassetto, Alessandro Torrielli and 
Federica Vian for constructive comments. 
This work was supported in part by the grant INTAS--00--390.
The work of J.A.\ and Y.M.\ is supported in part by
MaPhySto founded by the Danish National Research Foundation.
A.D.\ and Y.M.\ are partially supported by 
the Federal Program of the Russian Ministry of Industry,
Science and Technology No 40.052.1.1.1112.

%\setcounter{section}{0}
%\eop
\vspace{24pt}

\section*{Appendices}

\app{Derivation of the representation (\ref{1.3})\label{appA}}

To derive the path-integral representation (\ref{1.3}),
let us first observe that, applying the integral form (\ref{1.25}) of the
star-product (\ref{1.8}), we get for the star-product of an
{\em even}\/ number $M$ of functions:
\be
f _1 ({\bbox x}) \ast \ldots \ast f _M ({\bbox x})= \sum_{\xi_1,\ldots,\xi_M}
\e^{\frac {\i}2 (\theta^{-1})_{\mu\nu} \xi_l^\mu G^{-1}_{lj}\xi_j ^\nu }
f _1 ({\bbox x}+{\bbox \xi_1}) \cdots f _M ({\bbox x}+{\bbox \xi_M})\,,
\label{productM}
\ee
where the measure reads
\be
\sum_{\xi_1,\ldots,\xi_M}...~=~
\prod_{j=1}^{M}\frac{\d^{D}\xi^{\mu}_{j}}{\sqrt{\pi^D |\det \theta |}}...~,
\label{1.6}
\ee
while
\be
G^{-1}_{lj}=2(-1)^{l-j+1}\epsilon_{lj}~,~~~~~~~
G_{lj}=-\frac{1}{2}\epsilon_{lj}~,
\label{1.14}
\ee
with
\be
\epsilon _{lj}=
\left\{ 
\begin{array}{rc}
1& l<j \\
0& l=j \\
-1 & l>j \\
\end{array}\right.
\;~.   
\label{1.15}
\ee
This formula can be proved by induction.

Next, if some (even number) of $f$'s
equal 1, the Gaussian integral over the proper variables, which the
integrand does not depend on, can be performed reproducing Eq.~(\ref{productM})
for the lower number of nontrivial functions. Analogously, we get
\bea
\lefteqn{
f _1 ({\bbox x}) \ast \ldots \ast f _M ({\bbox x})\ast f _{M+1} ({\bbox x})~= 
\sum_{\xi_1,\ldots,\xi_M}
\e^{\frac \i2 (\theta^{-1})_{\mu\nu} \xi_l^\mu G^{-1}_{lj}\xi_j ^\nu } }\non
&&\times f _1 ({\bbox x}+{\bbox \xi_1}) \cdots f _M ({\bbox x}+{\bbox \xi_M})
f _{M+1} ({\bbox x}+\sum_{n=1}^M(-1)^n{\bbox \xi_n})
\label{productM+1}
\eea
for the star-product of an odd number of functions. Therefore, employing the
notation (\ref{1.4}) for the $\xi$-averaging,
Eq.~(\ref{productM}) can be rewritten as
\beq
\prod_n \ast f_n({\bbox x}) = \Big\langle \prod_n f_n ({\bbox x}
+{\bbox \xi_n}) \Big\rangle_\xi~.
\label{averageM}
\eeq
In particular, the simplest average is
\beq
\Big\langle  \xi^\mu_l \xi^\nu_j
\Big\rangle_\xi = - \frac \i2 \,\theta^{\mu\nu}\, \epsilon _{lj}
\label{propagator}
\eeq
since the inverse to the matrix $G^{-1}_{ij}$ is given by Eq.~(\ref{1.14}).
%In turn, eq. (\ref{propagator}) can be used for an alternative proof of
%eq. (\ref{averageM}). Altogether, it completes the formal proof of eq.
%(\ref{1.3}) once one identifies
%\be
%f _j ({\bf x})~\longrightarrow{~
%exp\Big[i\Delta y_{\mu}{\mathcal{A}}_{\mu}\left({\bf x}+{\bf y}(\tau_{j})
%\right)\Big]}~,
%\label{1.15a}
%\ee
%where the trajectory ${\bf x}(\tau)={\bf x}+{\bf y}(\tau)$ has been
%redefined in terms of ${\bf x}\equiv{\bf x}(0)$ and ${\bf y}(\tau)$ with
%${\bf x}(0)=0$.

Finally, let us note the following subtlety. 
If we make a finite-dimensional approximation of the functional space,  say, 
by means of the stepwise regularization, we
formally get from Eq.~(\ref{1.4}) the pattern of Eq.~(\ref{Gnaive}).
However, the measure in this case is not of the Wiener type, typical
trajectories are not continuous%
\footnote{This can be directly seen from the form of the matrix
$G^{-1}_{ij}$ given by Eq.~(\ref{1.14}), which is obviously nonlocal.} 
and uncertainties of the type $0\times \infty$ will appear 
when $\dot \xi$ is involved in calculations. % as we shall see in a moment. 
We shall rather keep $G$ smeared over an interval
$\varepsilon \sim 1/M$ to do the uncertainties,
while the results will be independent of the form of the smearing.
After doing the uncertainties we set $\epsilon=0$.

More complicated averages involving a functional $F[\xi]$
can be calculated using the Schwinger--Dyson equation
\beq
\Big\langle \xi^\mu(\tau)F[\xi] \Big\rangle_\xi =
\i \theta^{\mu\nu} \int \d\tau' \,G(\tau,\tau') 
\Bigg\langle \frac{\delta F[\xi]}{\delta\xi^\nu(\tau')}\Bigg\rangle_{\!\xi}\,,
\label{SD}
\eeq
which results from the invariance of ${\cal D} \xi$ under an infinitesimal
variation of the function $\xi^\mu(\tau)$. In particular, after the
substitution $F[\xi]=\xi^\nu(\tau')$, we obtain 
\beq
\Big\langle  \xi^\mu (\tau) \,\xi^\nu (\tau')
\Big\rangle_\xi =  \i \theta^{\mu\nu} G(\tau,\tau')~.
\label{Gpropagator}
\eeq
On the other hand, differentiating Eq.~(\ref{SD}) with respect to $\tau$, we
obtain another useful formula
\beq
\Big\langle \dot \xi^\mu(\tau)F[\xi] \Big\rangle_\xi =
\i \theta^{\mu\nu} \int \d\tau' \,\dot G(\tau,\tau') 
\Bigg\langle\frac{\delta F[\xi]}{\delta\xi^\nu(\tau')} \Bigg\rangle_{\!\xi}
\label{dotSD}
\eeq
which is to be employed when $\dot \xi$ enters the relevant averages.

Using this technique, Eq.~(\ref{1.3})
comes as a result of Feynman's disentangling of the
star-products and can be proved by expanding in the
powers of ${\cal A}$ and using Eq.~(\ref{Gpropagator}) whose
$\varepsilon\rightarrow0$ limit
is given by Eq.~(\ref{Gnaive}). There are no uncertainties 
at this level since $\dot \xi$ is not involved. In particular, this allows us 
to reduce Eq.~(\ref{Gpropagator}) to Eq.~(\ref{1.4a}).

To illustrate the subtleties with $\dot \xi$, let us calculate the variation
of $W(C)$ at an intermediate
point $\tau$ which should reproduce the
noncommutative field strength~\rf{defF}.
Applying the variational derivative to the right-hand side of \eq{1.3},
we get%
\footnote{An extra term $\i\A_\mu(x(\tau))\left(
\delta(\tau-\tau_f) -\delta(\tau-\tau_0)\right)$
emerges at the end points  as usual.}
\beq
\frac{\delta}{\delta x^\mu(\tau)}  W(C) =
\i \lim_{\varepsilon\ra0}\Big\langle 
\left( \left( \partial_{\mu} \A_\nu(\tau) -\partial_{\nu} \A_\mu(\tau)
\right)\dot x^\nu(\tau) - \partial_{\nu} \A_\mu(\tau) \dot \xi^\nu(\tau)\right)
\e^{\i \int \d   x^\rho \A_\rho \left( \bbox x+\bbox \xi \right)}
\Big\rangle_\xi  \,,
%=\lim_{\varepsilon\ra0}\Big\langle 
%\left( \left( \partial_{\mu} \A_\nu(\tau) -\partial_{\nu} \A_\mu(\tau)
%\right)\dot x^\nu(\tau) +  \theta^{\nu\lambda} \int d\sigma
%\partial_{\nu} \A_\mu(\tau) \dot G(\tau, \sigma)
%\partial_{\lambda} \A_\rho(\sigma) 
%\dot x^\rho(\sigma)\right)
%\e^{i \int d   x^\rho \A_\rho \left( x+\xi \right)}
%\Big\rangle_\xi 
\eeq
where we denoted $\A_\mu(\tau)\equiv \A_\mu(\bbox x(\tau)+\bbox \xi(\tau))$ 
for brevity.
Using \eq{dotSD} we can replace here $\dot \xi^\nu(\tau)$ by
\beq
\dot \xi^\nu(\tau) \stackrel{\rm w.s.}{=} - \theta^{\nu\lambda} \int \d\sigma
\,\dot G(\tau, \sigma)
\partial_{\lambda} \A_\rho(\sigma) 
\dot x^\rho(\sigma)\,,
\label{dotxi}
\eeq
which holds in the weak sense, \ie under the averaging over $\xi$.
This yields explicitly
\bea
{\cal F}_{\mu\nu}(\bbox x)&=& \lim_{\varepsilon\ra0}\Bigg\langle 
\Big(  \partial_{\mu} \A_\nu(\bbox x) 
-\partial_{\nu} \A_\mu(\bbox x)  \non
&&~~~~~~ + \theta^{\nu\lambda} 
 \int \d\sigma \,\dot G(\tau, \sigma)\partial_{\nu} 
\A_\mu(\bbox x+\bbox \xi(\tau)) 
\partial_{\lambda} \A_\rho(\bbox x+\bbox \xi(\sigma)) \Big)
\Bigg\rangle_\xi  \,.
\label{forF}
\eea
It is easy to see this is indeed a correct formula expanding in $\xi$
and using \eq{Gpropagator}. The combinatorics is as follows
\beq
\lim_{\varepsilon\ra0}
\frac{n!}{n!n!}\int \d\sigma \,\dot G(\tau, \sigma)G^n(\tau, \sigma) 
= \frac{1}{n!}\frac{1}{(n+1)} =\frac{1}{(n+1)!}\,.
\eeq
If we were substitute the limiting value $G_0(\tau- \sigma)$
given by \eq{Gnaive} when
$\dot G_0(\tau- \sigma)=\delta(\tau- \sigma)$ into the right-hand side
of \eq{forF} before averaging, we would rather get for the star-product only
the term of the first order in $\theta$ 
since $G_0(0)=0$. 

A lesson we have learned from this exercise is that whenever 
$\dot \xi(\tau)$ appears inside the averaging it should be substituted 
according to \eq{dotxi} rather than just by its 
\mbox{$\varepsilon\ra0$} limit which is given by
$- \theta^{\nu\lambda} \partial_{\lambda} \A_\rho(\tau) 
\dot x^\rho(\tau)$. Stated differently, the star-commutator of
two functions is represented in the integrand of the path integral by
\beq
f(\bbox x)\ast g(\bbox x)-g(\bbox x)\ast f(\bbox x)\ra
\i \theta^{\nu\lambda} 
 \int \d\sigma \,\dot G(\tau, \sigma)\partial_{\nu} f(\bbox x+\bbox \xi(\tau)) 
\partial_{\lambda} g(\bbox x+\bbox \xi(\sigma)) \,.
\label{itshould}
\eeq

An application of this formula is to demonstrate the star-gauge
covariance of the right-hand side of \eq{1.3} under the star-gauge
transformation
\beq
\delta_\alpha \A_\mu = \partial_\mu \alpha + \i 
\left( \alpha \ast \A_\mu -\A_\mu \ast \alpha \right) \,.
\label{stargauge}
\eeq
Using Eqs.~\rf{stargauge}, \rf{itshould}, \rf{dotxi}, we have
\bea
\delta_\alpha W(C) &=&
\lim_{\varepsilon\ra0}\i\Big\langle \int \d\tau \dot x^\mu(\tau)
\delta_\alpha \A_\mu ( \bbox x(\tau)+\bbox \xi(\tau))
\e^{\i \int \d   x^\rho \A_\rho \left( \bbox x+\bbox \xi \right)}
\Big\rangle_\xi  \non &=&
\lim_{\varepsilon\ra0}\i\Big\langle \int \d\tau \dot x^\mu(\tau)
\Big(\partial_\mu \alpha (\tau)
-  \theta^{\nu\lambda} 
 \int \d\sigma \dot G(\tau, \sigma)\partial_{\nu} \alpha(\tau) 
\partial_{\lambda} \A_\mu(\sigma)        \Big)
\e^{\i \int \d   x^\rho \A_\rho \left( \bbox x+\bbox \xi \right)}
\Big\rangle_\xi \nonumber \\&=&
\lim_{\varepsilon\ra0}\i\Big\langle \int \d\tau 
\left( \dot x^\mu(\tau)+ \dot \xi^\mu(\tau) \right)
\partial_\mu \alpha (\tau)
\e^{\i \int \d   x^\rho \A_\rho \left( \bbox x+\bbox \xi \right)}
\Big\rangle_\xi \nonumber \\
&=&\lim_{\varepsilon\ra0}\i\Big\langle \int \d\tau \frac \d{\d\tau}
\alpha (\tau)
\e^{\i \int \d   x^\rho \A_\rho \left( \bbox x+\xi \right)}
\Big\rangle_\xi \non
&=& \lim_{\varepsilon\ra0}\i\Big\langle \left(
\alpha (\tau_f) -\alpha (\tau_0)\right)
\e^{\i \int \d   x^\rho \A_\rho \left( \bbox x+\bbox \xi \right)}
\Big\rangle_\xi \non
&=& \i\left( \alpha (\bbox x(\tau_f)) \ast W(C) - 
W(C) \ast\alpha (\bbox x(\tau_0))\right)
\eea
as it should.

\app{Evaluation of integrals for the Gaussian regularization\label{appD}}

For the Gaussian regularization~\rf{deltaG}, we explicitly have from
\eq{DR}\footnote{Here 
$
{\rm Erf}(x)=\frac{2}{\sqrt{\pi}}\int_0^x \d z \e^{-z^2}
$
is the standard error function.}
\be
D^{\rm (R)}_{22}(X) =  g^2
%\Big( \sqrt{\pi}B-  |X_1| \int\limits_0^{|X_1|/\theta\mu}
%\d z \e^{-z^2} -\frac{\theta\mu}{2}\e^{-X_1^2/(\theta\mu)^2}\Big) 
%\frac{\e^{-X_2^2/(\theta\mu)^2}}{\pi \theta \mu} \non
%&= &g^2
\Big( B-  \frac12 |X_1|\, {\rm Erf}\left(\frac{|X_1|}{\theta\mu}\right)
 -\frac{\theta\mu}{2\sqrt{\pi}}\e^{-X_1^2/(\theta\mu)^2}\Big) 
\frac{\e^{-X_2^2/(\theta\mu)^2}}{\sqrt{\pi} \theta \mu} \,.
\label{regpro}
\ee
Differentiating \eq{regpro} with respect to $X_1$ we find
\be
\frac{\partial}{\partial X_1}D^{\rm (R)}_{22}(X)=
-\frac{g^2}2 {\rm sign} (X_1) \, {\rm Erf}\left(\frac{|X_1|}{\theta\mu}\right)
\frac{\e^{-X_2^2/(\theta\mu)^2}}{\sqrt{\pi} \theta \mu}\,.
\label{regsign}
\ee
Both~\rf{regpro} and \rf{regsign} are
 analytic in $X$ at $X=0$ and reproduce the nonregularized 
formulas~\rf{1.2} and \rf{1.2p} for $|X|\gg \theta\mu$.

Given this definition, the star-product on the left-hand side 
of \eq{f2} equals
\be
\delta^{(2)}_{\rm (R)}(X) * \delta^{(2)}_{\rm (R)}(Y)= \frac{1}{4\pi^2\theta^2}
 \e^{\i X\wedge Y /\theta -\mu^2 X^2/4- \mu^2 Y^2/4} \,.
\label{deltaregstar}
\ee
It is worth noting that the same expression can be obtained if we
do not smear the delta-functions and propagators but rather
modify the star-product by including the IR cutoff:
\be
\int \frac{\d^2 \xi\,\d^2 \eta}{4\pi^2\theta^2}
\e^{-\mu^2 \xi^2/4- \mu^2 \eta^2/4}\e^{\i \xi\wedge \eta /\theta }
\delta^{(2)}(X+\xi)\delta^{(2)}(Y+\eta)=
\frac{1}{4\pi^2\theta^2}
 \e^{\i X\wedge Y /\theta -\mu^2 X^2/4- \mu^2 Y^2/4}\,.
\ee
%For the star-product of the regularized propagators, we analogously find
%\bea
%\lefteqn
%{D_{22}^{\rm (R)}(X)* D_{22}^{\rm (R)}(Y)
%= 
%\frac{g^4}{4\pi^2\theta^2}
%\e^{\i X\wedge Y/\theta-\mu^2 X^2/4- \mu^2 Y^2/4}}\non
%&&\times\int\limits_{-\infty}^{+\infty} \d u 
%\e^{-\mu^2u^2/4 + u (\i Y_2/\theta-\mu^2 X_1/2)} \Big( B-\frac12 |u| \Big)
%\int\limits_{-\infty}^{+\infty} \d v \e^{-\mu^2v^2/4 - 
%v(\i  X_2/\theta+\mu^2 Y_1/2)} \Big( B-\frac12 |v| \Big)
%\non &&
%\label{starpropagatorG}
%\eea
%or
%\bea
%\lefteqn
%{D_{22}^{\rm (R)}(X)* D_{22}^{\rm (R)}(Y) = 
%\frac{g^4}{4\pi^2\theta^2} }\non
%&&\times\int\limits_{-\infty}^{+\infty} \d u 
%\e^{-\mu^2u^2/4 + \i u Y_2/\theta} \Big( B-\frac12 |u-X_1| \Big)
%\int\limits_{-\infty}^{+\infty} \d v \e^{-\mu^2v^2/4 - 
%\i v  X_2/\theta} \Big( B-\frac12 |v-Y_1| \Big)
%\non &&
%\label{starpropagatorGa}
%\eea
%which regularizes the ``naive'' 
%\eq{1.24} for $X_2,Y_2\lesssim \mu\theta$, reproducing the usual
%product of the two propagators as $\theta\ra0$.
For the star-product of the regularized propagators we analogously obtain
\eq{starpropagatorGa} which can be also rewritten as
\be
D_{22}^{\rm (R)}(X)* D_{22}^{\rm (R)}(Y)=
\int \frac{\d^2 \xi\,\d^2 \eta}{4\pi^2\theta^2}
\e^{-\mu^2 \xi^2/4- \mu^2 \eta^2/4}\e^{\i \xi\wedge \eta /\theta }
D_{22}^{\rm (0)}(X+\xi)D_{22}^{\rm (0)}(Y+\eta)
\ee
in analogy with \eq{deltaregstar}.

The right-hand side of \eq{starpropagatorGa} involves the integrals 
\be
I(X_1,Y_2)=-\frac{1}{4\pi\theta} 
\int\limits_{-\infty}^{+\infty} 
\d u \e^{-\mu^2u^2/4 + \i u Y_2/\theta} |u-X_1|\,.
\label{intG}
\ee
Differentiating with respect to $X_1$, as is needed for the 
observable~\rf{dumbbella} from Sect.~\ref{s:a.o.}, we get
\bea
\frac{\partial I(X_1,Y_2)}{\partial X_1}&=&\frac{1}{4\pi\theta} 
\int\limits_{-\infty}^{+\infty} \d u 
\e^{-\mu^2u^2/4 + \i u Y_2/\theta} {\rm sign}{(u-X_1)} \non
&=& \frac{1}{2\pi\theta}
\left(-\int\limits_0^{X_1} \d u \e^{-\mu^2u^2/4} \cos{\frac{uY_2}{\theta}}
+ \i\int\limits_{X_1}^\infty \d u \e^{-\mu^2u^2/4} \sin{\frac{uY_2}{\theta}}
\right) \nonumber \\
%&=& \frac{1}{2\pi\theta}
%\left(-\int\limits_0^{X_1} \d u \e^{-\mu^2u^2/4 + \i u Y_2/\theta}
%+ \i\int\limits_{0}^\infty \d u \e^{-\mu^2u^2/4} \sin{\frac{uY_2}{\theta}}
%\right) \non 
&=&
\frac{1}{2\pi Y_2}
\left(-\!\!\!\! \int\limits_0^{X_1 Y_2/\theta} \d \kappa 
\e^{-\nu^2 \kappa^2/4 + \i \kappa}
+ \i\int\limits_{0}^\infty \d \kappa \e^{-\nu^2\kappa^2/4} \sin{\kappa}
\right)
\label{intGsign}
\eea
where
\be
\nu=\frac{\mu \theta }{Y_2} \,.
\ee

As $\nu\ra0$ under normal circumstances, the first integral on the 
right-hand side yields
\be
-\frac{1}{2\pi Y_2}
\int\limits_0^{X_1 Y_2/\theta} \d \kappa 
\e^{\i \kappa} =
\i \frac{1}{2\pi Y_2} \left( \e^{\i X_1 Y_2/\theta }-1\right)
\ee
which itself would not give the anomaly since it vanishes as 
$\theta\ra\infty$. The second integral
\be 
\i\int\limits_{0}^\infty \d \kappa \e^{-\nu^2\kappa^2/4} \sin{\kappa}
= \frac{\sqrt{\pi}}{\nu} \e^{-1/\nu^2} {\rm Erf}\left(\frac{\i}{\nu}\right)
\ra \i
\ee
as $\nu\ra0$. This results in the anomalous behavior 
of the $1/\theta$-expansion.

%In the opposite limit of $\theta\ra0$

To calculate  the (regularized) diagrams 
in Figs.~\ref{fi:anobs}a and b, whose contributions are given by 
the left-hand sides of Eqs.~\rf{1)} and \rf{2)} 
with the propagators regularized according 
to \eq{regpro}, 
it is convenient to change the order of integration and first 
to integrate over the contour and
then over $\xi$ and $\eta$ representing the star-product. This will
be also convenient~\cite{ADM04} for higher-order calculations.

After the integration over $y_2$ and $z_2$, we have for the diagram 
in Fig.~\ref{fi:anobs}b
\bea
\lefteqn{\hbox{Fig.~\ref{fi:anobs}b}~=}\non
&&\frac{g^4}{4\pi^2}\int\limits_0^{\infty} \d \xi\int\limits_0^{\infty} 
\d \eta \e^{-\tilde\mu^2 \xi^2/4-\tilde\mu^2 \eta^2/4}
\left(\frac{(\cos \xi -1)}\xi\frac{(\cos \eta -1)}\eta
+\frac{\xi^2(\cos\eta-1)-\eta^2(\cos\xi-1)}{\xi\eta(\xi^2-\eta^2)}
\right) \non &&
\label{goodforsmall}
\eea
or
\be
\hbox{Fig.~\ref{fi:anobs}b}~=
\frac{g^4}{4\pi^2}\int\limits_0^{\infty} \d \xi\int\limits_0^{\infty} 
\d \eta \e^{-\tilde\mu^2 \xi^2/4-\tilde\mu^2 \eta^2/4}
\left(\frac{\cos \xi }\xi\frac{\cos \eta }\eta
+\frac{\eta^2\cos\eta-\xi^2\cos\xi}{\xi\eta(\xi^2-\eta^2)}
\right),
\label{goodforlarge}
\ee
where
\be
\tilde \mu = \frac{\mu\theta}R
\ee
and $R$ is the distance between the points along the axis 2 ($R=|x_2-y_2|$
for the diagram in Fig.~\ref{fi:anobs}b).
The representations \rf{goodforsmall} and \rf{goodforlarge} are
equivalent: the former formula 
is good for small $\xi$ and $\eta$ and the latter one
is good for large $\xi$ and $\eta$, where the integrals are manifestly
convergent.

The right-hand side of \eq{goodforlarge} (or \eq{goodforsmall})
can be identically represented for $|x_1-t_1|\ll 1/\mu$ as 
\bea
\hbox{Fig.~\ref{fi:anobs}b}&=&
\frac{g^4}{4\pi^2}\int\limits_0^{\infty} \frac{\d \xi}{\xi}
\int\limits_0^{{\xi}} \frac{\d \eta}{\eta}
\e^{-\tilde\mu^2 \xi^2/4-\tilde\mu^2 \eta^2/4}
\left( \cos(\xi-\eta)-\cos\xi \right) \non
&&+\frac{g^4}{4\pi^2}\int\limits_0^{\infty} \frac{\d \xi}{\xi}\cos\xi
\int\limits_0^{{\xi}} \frac{\d \eta}{\eta}
\left(\e^{-\tilde\mu^2 (\xi-\eta)^2/4-\tilde\mu^2 \eta^2/4}-
\e^{-\tilde\mu^2 \xi^2/4-\tilde\mu^2 \eta^2/4}
 \right) \nonumber \\
&&+\frac{g^4}{2\pi^2}\int\limits_0^{\infty} \frac{\d \xi}{\xi}\cos\xi
\int\limits_0^{{\xi}} \frac{\d \eta}{\eta}
\left(\e^{-\tilde\mu^2 \xi^2/2-\tilde\mu^2 \eta^2/4}-
\e^{-\tilde\mu^2 \xi^2/2+\tilde\mu^2 \eta^2/4}
 \right)  \non
&&+\frac{g^4}{2\pi^2}\int\limits_0^{\infty} \frac{\d \xi}{\xi}\cos\xi
\int\limits_{{\xi}}^\infty \frac{\d \eta}{\eta}
\left(\e^{-\tilde\mu^2 \xi^2/2-\tilde\mu^2 \eta^2/4}-
\e^{-\tilde\mu^2 \xi^2/4-\tilde\mu^2 \eta^2/4}
 \right).
\label{GfI}
\eea
The integrals in the second to fourth lines of this equation vanish
as $\tilde \mu\ra0$, while that in the first line gives 
\be
\hbox{Fig.~\ref{fi:anobs}b}~=
\frac{g^4}{4\pi^2} \frac{\pi^2}6 \,.
\label{fib}
\ee
The contribution of the diagram in Fig.~\ref{fi:anobs}c is the same.

After differentiating \eq{starpropagatorGa} twice with respect to $X_1$
and integrating over $y_2$ and $t_2$,
we find that the contribution of the 
diagram in Fig.~\ref{fi:anobs}a for $|X_1|\ll 1/\mu$ is relatively simple:
\be
\hbox{Fig.~\ref{fi:anobs}a}~=
\frac{g^4}{2\pi^2}\int\limits_0^{\infty} \d \xi\int\limits_0^{\infty} 
\d \eta \e^{-\tilde\mu^2 \xi^2/4-\tilde\mu^2 \eta^2/4}
\delta^{(1)}(\xi) \cos\eta 
\left(\frac{RB}{\theta}-\frac12 \eta 
\right),
\label{relsimple}
\ee
which can be easily integrated 
as $\tilde \mu\ra0$ to give
\be
\hbox{Fig.~\ref{fi:anobs}a}~=\frac{g^4}{4\pi^2} \,.
\label{rl2}
\ee

Summing the contributions of all three diagrams in Fig.~\ref{fi:anobs},
we get \eq{notatall}
which was obtained in Sect.~\ref{s:a.o.}
by using the ``naive'' formula~\rf{1.24}. 
We have thus shown that it is reproduced for the Gaussian regularization.
%%The above calculation shows that it remains valid as $\theta\ra0$
%%at fixed $\mu$, when  $\tilde \mu\ra 0$ as well.

\app{Regularization by a box\label{someu}}

As the star-product of two propagators is regularization-dependent,
one may wonder what happens for other regularizations.
In this Appendix we introduce the regularization by putting the theory
in a box and discuss how the ``naive'' \eq{1.24}, which leads us to 
the nonvanishing $\theta^0$-term, is reproduced.
We also speculate how our 
results can be made compatible with previously obtained
results, in particular the results of the TEK model,
 %\be
%D_{22}(X)\ast D_{22}(Y)= 
%\frac{\theta^2}{4\pi^2}\frac{\e^{\i (X_1Y_2- X_2Y_1)/\theta}}{X_2^2 Y_2^2}\,.
%\label{starpropagatorn}
%\ee

Let us introduce an IR regularization by putting the theory in a box
of size $L_1\times L_2$. 
The star-product in the coordinate space equals
\bea
\lefteqn{D_{22}^{\rm (R)}(X)\ast D_{22}^{\rm (R)}(Y)= 
 \int\limits_V \frac{\d^2 \xi\,\d^2 \eta }{4\pi^2 \theta^2}
\e^{\i \xi \wedge \eta /\theta }D_{22}(X-\xi) D_{22}(Y-\eta)}
\non
&=& 
 \frac{g^4}{4\pi^2 \theta^2}
\int\limits_{-L_1/2}^{+L_1/2} \d \xi_1 \e^{\i \xi_1 Y_2} 
\Big(B-\frac12{|X_1-\xi_1|}\Big)
\int\limits_{-L_1/2}^{+L_1/2} \d \eta_1 
\e^{-\i \eta_1 X_2} \Big(B-\frac12 {|Y_1-\eta_1|}\Big) \nonumber \\ 
&=& \frac{g^4 \theta^2 }{4\pi^2}
\frac{\left[ \e^{\i X_1Y_2/\theta}-\Big(1+\i \frac{X_1Y_2}{\theta}\Big)
\cos\frac{L_1 Y_2}{2\theta} +\left(2B-\frac{L_1}{2}\right)  
\frac{Y_2}{\theta} 
\sin \frac{L_1 Y_2}{2\theta} 
\right]}{Y_2^2} \non
&& ~~~ \times
\frac{\left[ \e^{-\i X_2Y_1/\theta}-\Big(1-\i \frac{X_2Y_1}{\theta}\Big)
\cos\frac{L_1 X_2}{2\theta}+ \left(2B-\frac{L_1}{2}\right)  
\frac{X_2}{\theta} 
\sin \frac{L_1 X_2}{2\theta} 
\right]}{X_2^2}.  \non & &
\label{starpropagatorL}
\eea

Equation~\rf{starpropagatorL} differs from \rf{1.24} by terms  
at most linear in either $X_1$ or $Y_1$. These terms vanish
when applying $\partial ^2/\partial X_1^2$ or 
$\partial ^2/\partial Y_1^2$ and we obtain  again \eq{1.24ff}:
\be
\delta^{(2)}(X)\ast \delta^{(2)}(Y)= \frac{1}{4\pi^2 \theta^2}
\e^{\i X\wedge Y /\theta} \,.
\ee

If $L_1\ra\infty$ at fixed $\theta$, the extra terms in 
\rf{starpropagatorL} oscillate strongly
and can most probably can be omitted, reproducing \eq{1.24}.
This would be similar to how the one-dimensional propagator
\be
D(p)=\int\limits^{+L/2}_{-L/2} \d 
x \left(B -\frac12 |x|  \right) \e^{\i px} 
=\frac{1}{p^2} \left(1-\cos\frac{Lp}2\right)
+\frac 1p\left(2B-\frac{L}2 \right)\sin\frac{Lp}2
\label{ppp}
\ee
reproduces $1/p^2$ when $Lp\gg1$, which means that typical distances are
much smaller than $L$. Note that nothing depends 
on the constant $B$ for such distances, while the formulas of this 
Section are simplified for $B=L_1/4$.

Alternatively, if $\theta\ra\infty$ at fixed $L_1$ (like on a torus),
$L_1 /{\theta}\ra 0$ and we obtain from \eq{starpropagatorL}
\be
D_{22}(X)\ast D_{22}(Y)= 
\frac{g^4\theta^2}{4\pi^2}
\frac{\left( \e^{\i X_1Y_2/\theta}-1-\i \frac{X_1Y_2}{\theta}\right)}{Y_2^2} 
\frac{\left(\e^{-\i X_2 Y_1/\theta}-1+\i \frac{X_2Y_1}{\theta}\right)}{X_2^2},
\label{starpropagator1}
\ee
and the expansion in $1/\theta$ begins with a term $\propto \theta^{-2}$.
Therefore, the terms of the order $\theta^0$ and $\theta^{-1}$ would 
vanish.

On a torus with periods $L_1$ and $L_2$ we have
\be
\theta = \frac{L_1 L_2}{2\pi} \Theta \,,
\label{Theta}
\ee
where (the irrational) $\Theta$ is the dimensionless noncommutativity 
parameter. Using \eq{Theta} we find
\be
\frac{L_1 X_2}{2\theta} = \frac{\pi}{\Theta} \frac {X_2}{L_2} 
\label{arg}
\ee
which is small for $\Theta \ga 1$ and $X_2 \ll L_2$, 
\ie for the case of loops much
smaller than the period, as is required for an
approximation of ${\Bbb R}^2$ by ${\Bbb T}^2$.
Then \eq{starpropagator1} is reproduced. 
One might also expect than ${L_1 X_2}/{2\theta}$ is a multiple of $2\pi$
for the regularization by a {\it discrete} torus, so then \eq{starpropagator1}
is exact, thereby explaining the relation with genus expansion in TEK. 

Remarkably, \eq{starpropagator1} (or even more general \eq{starpropagatorL})
is consistent with the NC loop equation. Applying the loop operator
to the nonplanar diagram of the order $g^4$ involving 
the star-product~\rf{starpropagatorL},
we get for the IR regularization by a box:
\be
-\frac{g^4}{4\pi^2}V \oint\limits_C \d z_\nu
\int\limits_{C_{xz}} \d y_2 
\int\limits_{C_{zx}} \d t_2 \frac{\e^{\i X_1Y_2 /\theta}
\left[ \e^{-\i X_2Y_1/\theta}-\Big(1-\i \frac{X_2Y_1}{\theta}\Big)
\cos\frac{L_1 X_2}{2\theta} +\Big( 2B-\frac{L_1}{2}\Big)
\frac{X_2}{\theta} 
\sin \frac{L_1 X_2}{2\theta} 
\right]}{X_2^2}\,.
\label{RRhs}
\ee
For the right-hand side of the loop equation we 
use \eq{4RHS}, which has in  the first line the same
integrals as in \eq{starpropagatorL}. This results in the same
expression as \rf{RRhs}, which replaces the second line of \rf{4RHS}.

It is worth noting that the modification \rf{starpropagator1} does not cure 
the shape-dependence of the $\theta^{-2}$-term. It now equals
\bea
\theta^{-2} \hbox{-term}~&=&\frac{g^4}{96\pi^2\theta^2 }
\left( 1+\frac{5}{\pi^2} \right) A^4 
\qquad \hbox{for circle}
\non\theta^{-2} \hbox{-term}~&=&\frac{g^4}{96\pi^2\theta^2 }
\cdot \frac 32 A^4  
\qquad \hbox{for rectangle}\, ,
\eea
and we still seem to have an explicit breaking of 
symplectic invariance in the $1/\theta$-expansion.

%\eop

\end{document}